\documentclass[linenumbers, twocolumn, tighten]{aastex631}
\nolinenumbers

\usepackage{xspace}

\usepackage{glossaries}
\setacronymstyle{long-short}
\glsdisablehyper

\usepackage{CJKutf8}
\newcommand{\JOname}{{\begin{CJK}{UTF8}{gbsn}(王加冕)\end{CJK}}}

\usepackage{newunicodechar}
\DeclareRobustCommand{\okina}{%
  \raisebox{\dimexpr\fontcharht\font`A-\height}{%
    \scalebox{0.8}{`}%
  }%
}
\newunicodechar{ʻ}{\okina}

\graphicspath{{./}{figures/}}
\usepackage{amsmath,amssymb,mathptmx,txfonts,tikz,bm}
\usepackage[capitalize]{cleveref}
\usepackage{xspace}

\newcommand{\annotate}[2]{\begin{tikzpicture}
    \node[anchor=south west,inner sep=0,align=center] (image) at (0,0) {
    #1
    };
    \begin{scope}[x={(image.south east)},y={(image.north west)}]
    #2
    \end{scope}
\end{tikzpicture}}

\def\bibitem{\vskip\bibskip\footnotesize\savebibitem}

\newacronym{PDF}{PDF}{probability density function}

\newacronym{PCA}{PCA}{principal component analysis}

\newacronym{PPF}{PPF}{probability point function}

\newacronym{CDF}{CDF}{cumulative distribution function}

\newacronym{TESS}{TESS}{Transit Exoplanet Survey Satellite}
\newcommand{\tess}{\gls{TESS}\xspace}

\newacronym{SPOC}{SPOC}{Science Processing Operations Center}

\newacronym{TIC}{TIC}{TESS Input Catalog}

\newacronym{HR}{HR}{Hertzsprung-Russell}

\newacronym{FP}{FP}{false positive}

\newacronym{TP}{TP}{true positive}

\newacronym{PDS}{PDS}{power density spectrum}
\newcommand{\PDS}{\gls{PDS}\xspace}

\newacronym{PSD}{PSD}{power spectral density}

\newacronym{LC}{LC}{long cadence}

\newacronym{SC}{SC}{short cadence}

\newacronym{ACF}{ACF}{autocorrelation function}

\newacronym{ATL}{ATL}{Asteroseismic Target List}

\newacronym{RG}{RG}{red giant}
\newcommand{\RG}{\gls{RG}\xspace}

\newacronym{MS}{MS}{main-sequence}
\newcommand{\MS}{\gls{MS}\xspace}

\newacronym{SG}{SG}{subgiant}
\newcommand{\SG}{\gls{SG}\xspace}

\newacronym{FFI}{FFI}{full-frame image}

\newacronym{DR}{DR}{dimensionality reduction}

\newacronym{KDE}{KDE}{kernel density estimate}

\newacronym{MCMC}{MCMC}{Markov Chain Monte Carlo}

\newacronym{S/N}{S/N}{signal-to-noise ratio}
\newcommand{\snr}{\gls{S/N}\xspace}

 \newcommand{\pmodes}{p-modes\xspace}
\newcommand{\pmode}{p-mode\xspace}
\newcommand{\gmodes}{g-modes\xspace}
\newcommand{\gmode}{g-mode\xspace}
 \newcommand{\kepler}{\textit{Kepler}\xspace}

\newcommand{\muHz}{\,\mu\mathrm{Hz}}

\newcommand{\K}{\,\mathrm{K}}

\newcommand{\Msun}{\,\mathrm{M}_\odot}

\newcommand{\numax}{\nu_{\mathrm{max}}\xspace}
\newcommand{\dnu}{\Delta\nu\xspace}

\newcommand{\DPione}{\Delta\Pi_1\xspace}

\newcommand{\dotwo}{\delta\nu_{02}}

\newcommand{\Teff}{\,T_\mathrm{eff}}
\newcommand{\eps}{\,\varepsilon}
\newcommand{\bprp}{\,G_{\mathrm{BP}}-G_{\mathrm{RP}}}

\newcommand{\pbjam}{\texttt{PBjam}\xspace}

\definecolor{ForestGreen}{HTML}{228B22}

\begin{document}

\title{Asteroseismology with PBjam 2.0: measuring dipole mode frequencies in coupling regimes from main sequence to low-luminosity red giant stars.}

\correspondingauthor{M. B. Nielsen}
\email{m.b.nielsen.1@bham.ac.uk}

\author[0000-0001-9169-2599]{M. B. Nielsen}
\affiliation{School of Physics and Astronomy, University of Birmingham, Birmingham B15 2TT, UK}

\author[0000-0001-7664-648X]{J. M. J. Ong \JOname}
\altaffiliation{NASA Hubble Fellow}
\affiliation{Institute for Astronomy, University of Hawaiʻi at Mānoa, 2680 Woodlawn Drive, Honolulu, HI 96822, USA}

\author[0000-0002-1389-1549]{E. J. Hatt}
\affiliation{School of Physics and Astronomy, University of Birmingham, Birmingham B15 2TT, UK}

\author[0000-0002-4290-7351]{G. R. Davies}
\affiliation{School of Physics and Astronomy, University of Birmingham, Birmingham B15 2TT, UK}

\author[0000-0002-5714-8618]{W. J. Chaplin}
\affiliation{School of Physics and Astronomy, University of Birmingham, Birmingham B15 2TT, UK}

\author[0009-0002-8134-4026]{G. T. Hookway}
\affiliation{School of Physics and Astronomy, University of Birmingham, Birmingham B15 2TT, UK}

\author[0000-0002-5496-365X]{A. Stokholm}
\affiliation{School of Physics and Astronomy, University of Birmingham, Birmingham B15 2TT, UK}

\author[0000-0002-7603-3525]{O. J. Scutt}
\affiliation{School of Physics and Astronomy, University of Birmingham, Birmingham B15 2TT, UK}

\author[0000-0001-9214-5642]{M. N. Lund} 
\affiliation{Stellar Astrophysics Centre, Department of Physics and Astronomy, Aarhus University, Ny Munkegade 120, DK-8000 Aarhus C, DK}

\author[0000-0002-8854-3776]{R. A. García} 
\affiliation{Université Paris-Saclay, Université Paris Cité, CEA, CNRS, AIM, 91191, Gif-sur-Yvette, France}

\begin{abstract}
\nolinenumbers

\pbjam is an open-source software package for measuring mode frequencies of solar-like oscillators. These frequencies help constrain stellar evolution models to precisely estimate masses, radii, and ages of stars. The overall aim of \pbjam is to simplify this process to the point where it may be done by non-experts or performed on thousands of stars with minimal interaction. The initial release of \pbjam was restricted to only identifying modes of $\ell=0$ and $\ell=2$, since these are the simplest to treat consistently across different stellar evolutionary stages. Here we introduce a new set of three separate models which lets \pbjam automatically identify $\ell=1$ modes in stars that experience varying degrees of coupling between p- and \gmodes. These include a simple asymptotic relation for \pmodes which can be applied to main-sequence stars, a matrix formalism aimed at treating frequency dependent coupling in sub-giants, and a uniform coupling model which is suitable for red giants. These models follow the Bayesian methodology established in the first release of \pbjam, where a large set of previous observations is used to construct a nonparametric prior probability density for the new set of model parameters. This extension allows \pbjam to build a more complete description of the power due to oscillations across a wider range of evolutionary stages.  
\end{abstract}

\keywords{Asteroseismology (73) --- Astronomy data analysis (1858) --- Open source software (1866)}

\section{Introduction} 
\label{sec:intro}
Stars are able to oscillate in a variety of different ways \citep{Aerts2021}, and using these oscillations to constrain their interior physics is an increasingly important part of characterizing stellar systems. Stars with thick outer convective envelopes, like the Sun, oscillate primarily due to turbulent stresses generated by the convection. Such solar-like oscillations are both driven and subsequently damped by convective motion in the star \citep{Chaplin2013}, and are commonly observed in main sequence stars with masses less than $\approx1.6\Msun$, as well as those that have evolved to become red giants \citep[see, e.g.,][]{Garcia2019}. The identification of a star as a solar-like oscillator may be made either visually or automatically based on the \PDS of time series of radial velocity \citep{Kjeldsen2005, Campante2024} or more often from photometric flux measurements \citep{Verner2011, Hon2021, Hatt2023}. Solar-like oscillations appear in the \PDS as a roughly Gaussian-shaped envelope of excess power centered on a characteristic frequency $\numax$. This excess consists of contributions from many normal modes, each of which has a spherical harmonic angular degree, $\ell$, and azimuthal order, $m$. Several overtones of these modes may be visible, which are distinguished by their radial order $n$. For \MS stars, solar-like oscillations are pressure-dominated (\pmodes) and follow a regular pattern in frequency \citep{Tassoul1980}, where modes of the same angular degree but consecutive radial orders are separated by a near-constant spacing, $\dnu$. 

These global asteroseismic parameters, $\dnu$ and $\numax$, can be measured for thousands of stars \citep[e.g.,][]{Yu2018, Pinsonneault2025}, and provide a way to estimate stellar masses and radii to a precision of a few percent \citep{Kjeldsen1995,Stello2009, Miglio2021}. However, if the modes of the oscillations can be identified by their radial and angular degree and their frequencies measured accurately, the stellar parameters can be constrained to an even greater precision than by using the global parameters alone \citep{Lebreton2014, Stokholm2019, Cunha2021}. Since these oscillations provide precise stellar masses, radii, and ages, they allow these stars to be used as probes of Galactic evolution \citep{Miglio2013, Borre2022}, to characterize potential orbiting planets or companions \citep{Davies2015, Lundkvist2018, Huber2019}, and to study the physics and evolution of the stars themselves \citep{Lund2017, SilvaAguirre2017, Buchele2024}. 

\begin{figure*}
    \centering
    \includegraphics[width=\linewidth]{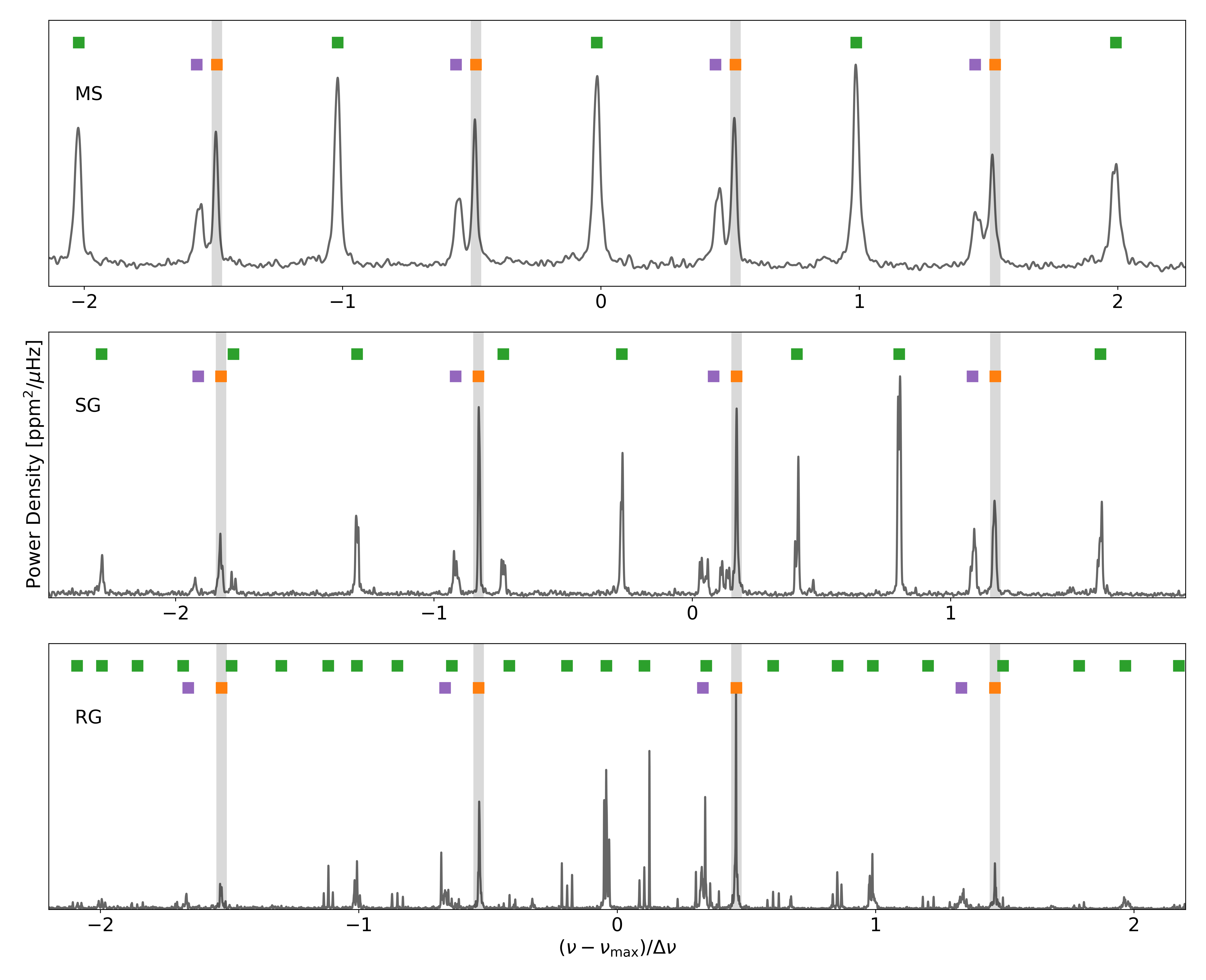}
    \caption{Example oscillation spectra of stars at three different evolutionary stages. The frequency axis is centered on $\numax$, and is shown in units of $\dnu$. The power density is shown in dark gray, and the modes are labeled according to their angular degree, where modes of $\ell=0$, $\ell=1$, and $\ell=2$ are shown as orange, green, and purple squares respectively. The $\ell=0$ modes, highlighted by the shaded regions, have been aligned between the three frames for clearer comparison. Top: A spectrum of the \MS star KIC~5184732 where all modes appear with a regular pattern. Middle: A spectrum of the \SG star KIC~5723165 where the dipole modes begin to deviate from the otherwise regular pattern of the $\ell=2,0$ mode pairs. Bottom: A spectrum of the \RG star KIC~4448777 where the regular pattern of the $\ell=1$ modes has almost completely disappeared.}
    \label{fig:examplespectra}
\end{figure*}

\pbjam\footnote{\url{https://github.com/grd349/PBjam}} is an open source \texttt{Python} library which is meant to provide a set of tools to simplify or even automate the identification and characterization of these modes of oscillation in solar-like oscillators. The initial release of \pbjam \citep{Nielsen2021} was restricted in scope to the consideration of pure p-modes. In \MS stars, provided the oscillations have a high enough \snr, the labeling of such regularly-spaced p-modes is fairly straightforward. However, as stars evolve off the main sequence, pressure and gravity dominated modes of the same angular degree and azimuthal order can couple to each other when the regions in which they propagate inside a star are close to each other or overlap in terms of the stellar radius \citep[see, for example,][]{Osaki1975, Aizenman1977, Takata2016}. For a $1\Msun$ \MS star, the \gmode cavities are restricted to the internal radiative region, far below the stellar surface. The oscillations we can observe are therefore pressure-dominated and follow a repeating pattern like that shown in the top frame of Fig. \ref{fig:examplespectra}. However, as the star evolves off the main sequence and becomes a \SG, the inner boundary of the \pmode cavity reaches further into the star, while the \gmode cavity expands outwards. Observationally, the mode frequency of the visible \pmodes decreases, at which point the lowest-frequency \pmodes in the oscillation envelope may begin to couple with the highest-frequency \gmodes. 

Gravity waves (\gmodes) propagate horizontally, so \gmode oscillations cannot exist at $\ell=0$ \citep{Aerts2010}; as such, the corresponding radial \pmodes remain unaffected by them throughout the star's life. The same is not true for angular degrees $\ell > 0$ however, and for $\ell=1$ modes in particular the otherwise regular pattern of the modes begins to deviate from that seen on the main sequence as the modes start exhibiting mixed character (see the middle frame of Fig. \ref{fig:examplespectra}). Modes of $\ell=2$ can also couple, but the size of the evanescent regions separating the radiative core from the outer p-mode cavity increases with $\ell$, rendering the coupling between the two mode cavities increasingly weak. This means that $\ell\ge2$ modes of mixed character are only rarely observed (typically only when p- and g-modes are almost in resonance). 

Thus, off the main sequence, the $\ell=0$ and $\ell=2$ modes largely retain the simple pattern associated with pure p-modes, and they are therefore the easiest to identify throughout the evolutionary stages of solar-like oscillators. The features of dipole modes ($\ell=1$), however, rapidly become more complicated. The $\ell=1$ \pmodes start to couple with gravity-dominated modes that are otherwise hidden in the stellar interior, and the resulting mixed mode frequencies no longer follow a simple pattern that is easily identifiable by eye, much less in an automated fashion \citep{Li2020b, Appourchaux2020}. As the star evolves further to become a \RG star, multiple $\ell=1$ \gmodes may begin to couple to the same \pmode, at which point the mode pattern becomes much more complicated to identify (see the lower frame of Fig. \ref{fig:examplespectra}). 

In this paper, we describe an update to \pbjam, intended to solve this problem of identifying modes in a complex spectrum of oscillations. By doing so, we expand its capabilities to automate constraining the internal structure and rotation of evolved stars. We take a similar approach to that in the original release of \pbjam, defining a model of the power density spectrum consisting of a predetermined number of radial orders and a fixed range of angular degrees of $\ell=0,\,1$ and $2$. The remaining parameters of the model that determine where those modes are located in frequency are left as free variables to be sampled subject to a prior probability density. This approach assumes that there is an underlying function that determines where modes are located in frequency. For the pure \pmodes, for example, this function is the asymptotic relation \citep{Tassoul1980} which produces the regularly spaced set of modes given a set of parameters which include $\numax$ and $\dnu$. For mixed modes, however, this is a set of functions that include the pure \gmode frequencies and their ability to couple to the \pmodes. 

The prior probability densities for these models are constructed based on a previously measured set of parameters for thousands of stars. Given a smaller set of observational inputs for a target star, \pbjam automatically constructs a prior distribution for a given model which is localized around the target in the parameter space. This local prior volume is then sampled given an observed spectrum of the star to estimate the posterior probability distribution of each  mode frequency. This method was used in the initial release of \pbjam but only included the ability to automatically identify the $\ell=2,0$ mode pairs. In this release of \pbjam, we extended this approach to include three separate methods to assign identifications and locations to $\ell=1$ modes in a similar automated fashion. These models consider cases of no coupling, coupling of a few \gmodes to many \pmodes, and many \gmodes to many \pmodes, and as such are broadly aimed at, but are not restricted to, being applied to \MS, \SG and low-luminosity \RG stars respectively.

Once the oscillation modes have been identified, the guess for the mode locations is passed on to a more detailed spectrum model, where many of the constraints applied in the mode identification stage are released. The spectrum model parameters in this stage are set at independent variables, which allows the detailed `peakbagging' model to more precisely capture effects that are not included in the mode identification models, such as stellar rotation and acoustic glitches, as well as any correlations that may be present between the parameters of adjacent modes. 

In Section \ref{sec:modeID} we describe the spectrum models that are used to perform the mode identification. These include a background model, a model consisting of just the $\ell=2,0$ modes, and three separate models for treating the $\ell=1$ modes. This section also discusses the treatment of rotation which requires special consideration for \RG stars. This is followed in Section \ref{sec:peakbag} by a description of the method used to perform the final detailed peakbagging, using the mode identification results from Section \ref{sec:modeID}. Finally in Section \ref{sec:conclusion} we summarize the updated capabilities of the \pbjam software package and possible extensions. 

\section{Mode identification}
\label{sec:modeID}
\begin{figure*}
    \centering
    \includegraphics[width=1\linewidth]{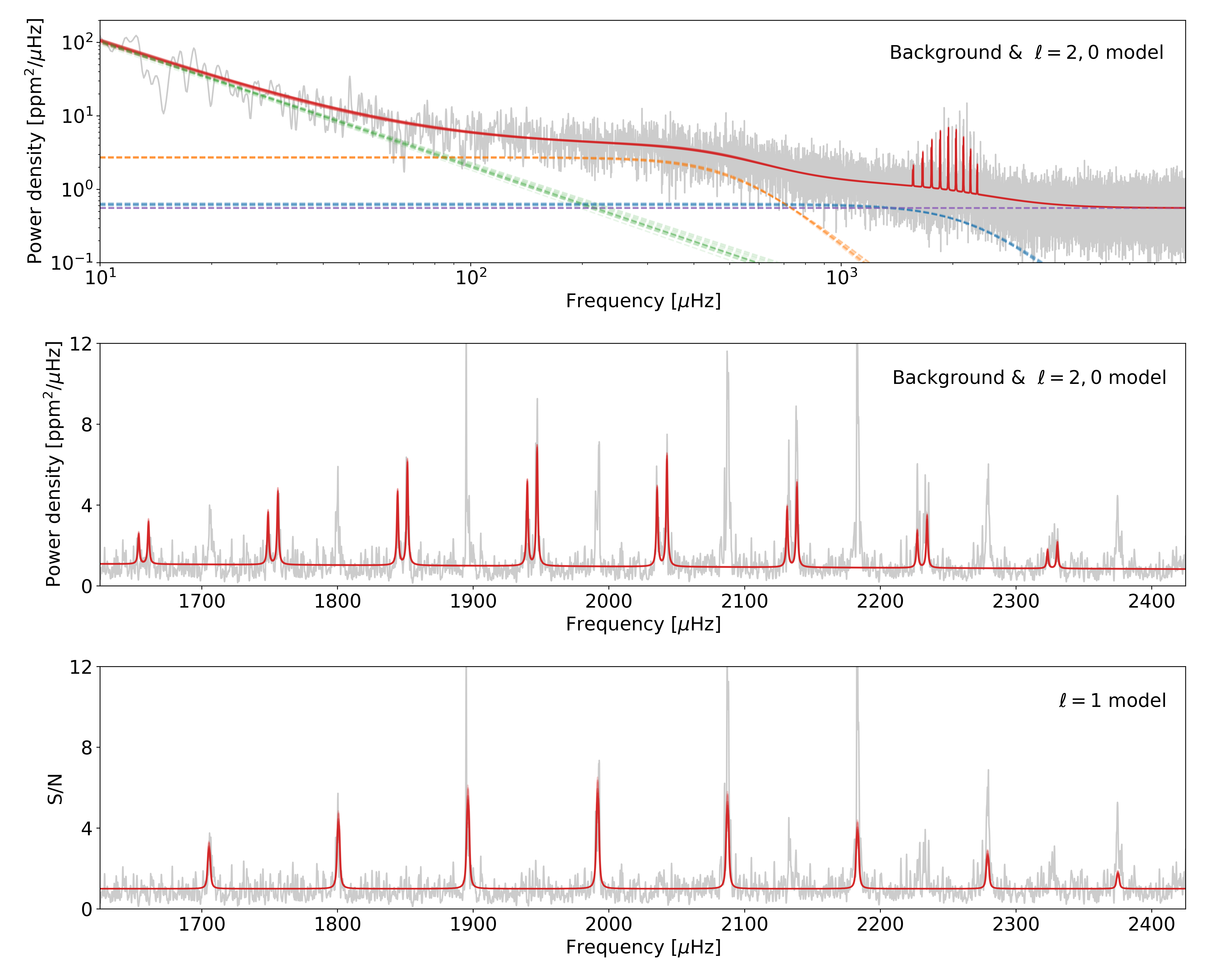}
    \caption{Models used in \pbjam for the \MS star KIC~5184732. The top frame shows the smoothed \PDS (gray) and a set of models (red) constructed from samples of the posterior distribution resulting from sampling Eq.~\ref{eq:background} and Eq.~\ref{eq:asymptotic_p_l0} which constitutes the first stage of the mode identification in \pbjam. The first, second, and third Harvey-like profiles are shown in dashed blue, orange, and green respectively, and the shot noise level is shown in dashed purple. The middle frame shows a smaller range around the oscillation envelope of the same model as in the top frame, for clarity. The lower frame shows the residual \snr spectrum in gray after dividing the \PDS by the median background and $\ell=2,0$ models, along with similarly drawn samples of the $\ell=1$ model posterior shown in red.}
    \label{fig:modelSteps}
\end{figure*}



With the addition of the $\ell=1$ mode identification, this part of \pbjam is now done in two steps. The first step is identifying the $\ell=0$ and $\ell=2$ modes, which is done by sampling the parameters of a model consisting of four background terms and a set of $\ell=2,0$ mode pairs. By dividing the \PDS by this model we obtain a residual spectrum which to a large extent only contains the $\ell=1$ modes, and a separate model is then defined for this residual \snr spectrum to identify their locations in frequency. 

This methodology was chosen since accurately sampling the parameter spaces separately in the two steps is more computationally tractable. In the following we describe the set of models used in the mode identification in more detail, and a complete list of parameters is shown in Table \ref{tab:modeIDvariables}.

\pbjam does not currently attempt to identify the location of $\ell=3$ modes as these are only rarely observed due to their low visibility in photometric observations \citep[see][]{Handberg2011}. They are therefore not expected to bias the result by, for example, being misidentified as $\ell=1$ modes, as the few $\ell=3$ modes that may be observable represent only a small fraction of the power compared to the more numerous $\ell=1$ modes. A model solution with the correct identification of the $\ell=1$ would therefore be more likely than one which proposes a misidentification.

\subsection{Identification of $\ell=2,0$ modes}
The fundamental part of \pbjam is the identification of the $\ell=2,0$ mode pairs. This allows us to precisely determine the global parameters $\dnu$ and $\numax$, as well as the approximate frequencies, heights, and widths for the $\ell=2,0$ modes. These are subsequently used in identifying the $\ell=1$ modes and the detailed peakbagging. 

To a large extent we use the same model and parameterization as in the original release of \pbjam. However, the original release only considered the \snr spectrum after removing an empirical estimate of the background noise. Here we construct a model of the \PDS which also includes background noise terms due to granulation and longer period variability like magnetic activity and instrumental effects. This model is given by 
\begin{equation}
    M(\nu) = W + \eta^2\,\left[B(\nu) + M_{\ell=2,0}\left(\nu\right)\right]
\end{equation}
where $W$ is the frequency independent shot noise level, $B\left(\nu\right)$ describes the frequency dependent background noise, and $M_{\ell=2,0}\left(\nu\right)$ describes the power due to the oscillations. The factor $\eta^2$ modulates the frequency-dependent terms to account for observing an otherwise continuous intensity variation at discrete times, and is given by $\eta^2=\mathrm{sinc}^2(\frac{\pi}{2}\frac{\nu}{\nu_{\mathrm{Nyquist}}})$, where $\nu_{\mathrm{Nyquist}}$ is the Nyquist frequency of the observations. We sample the parameters of the different terms (described in detail below) in the model simultaneously. 

\subsubsection{The background noise model}
Following \citet{Kallinger2014} the background model consists of three frequency-dependent Harvey-like profiles \citep{Harvey1985} and is given by
\begin{equation}
B\left(\nu\right) =\frac{a_{\mathrm{I}} / b_{\mathrm{I}}}{1 + \left(\nu/b_{\mathrm{I}}\right)^{c_{\mathrm{I}}}} + \sum\limits_{k=1}^{2} \frac{a / b_{k}}{1 + \left(\nu/b_k\right)^{c_k}}.
\label{eq:background}
\end{equation}
Here the first term is a Harvey-like profile with a characteristic frequency $b_I \sim 1\muHz$. This term describes variability which can be caused by, for example, magnetic activity and rotation but is often also due to residual long-term instrumental variability. This Harvey-like profile is characterized by the power, $a_I$, at $\nu=0$, and the exponent $c_I$ which governs the decrease of power with frequency. 

In addition to the instrumental terms we also include two Harvey-like profiles to describe the power due to convection on the stellar surface. This variability often appears with two characteristic frequencies, $b_k$, where the first ($k=1$) is comparable to $\numax$, and the second ($k=2$) shows significant power at frequencies below $\approx 0.3 \numax$. \citet{Kallinger2014} showed that the power in these two terms is often comparable, and so we use a single variable, $a$, to describe the power in each term. The top frame of Fig.~\ref{fig:modelSteps} shows a combination of the background model and its individual terms in relation to the $\ell=2,0$ model for the star KIC~5184732. 

\subsubsection{The $\ell=2,0$ model}
The model for the oscillations in the spectrum consists of a sum of Lorentzian profiles \citep{Anderson1990}, and is given as
\begin{equation}
M_{\ell=2,0}\left(\nu\right) = \sum\limits_{n_\mathrm{p}={n_{\mathrm{max}} - N_\mathrm{p}/2}}^{n_{\mathrm{max}} + N_{\mathrm{p}}/2}\sum\limits_{\ell=0,2}\sum\limits_{m=-\ell}^{\ell}\frac{H_{n_{\mathrm{p}} \ell m} }{1 + \frac{4}{\Gamma^2}(\nu-\nu_{n_\mathrm{p} \ell m})^2}.
\label{eq:l20mod}
\end{equation}
The first sum is over the \pmode radial orders, $n_\mathrm{p}$, the second sum is over angular degrees $\ell=0$ and $\ell=2$, and the final sum is over the $2\ell+1$ azimuthal orders, $m$, associated with each angular degree. Each Lorentzian profile is centered on a mode frequency $\nu_{n_\mathrm{p} \ell m}=\nu_{n_\mathrm{p} \ell} + m \delta\nu_{\mathrm{env}}$, such that a given $\ell=2,0$ pair consists of a single mode of $\ell=0$ and a multiplet of five $\ell=2$ modes which are evenly split in frequency by $\delta\nu_{\mathrm{env}}$. This parametrizes the effect the stellar rotation on the $\ell=2$ modes in the model, and we leave $\delta\nu_{\mathrm{env}}$ as a free variable which is assumed to be the same for all $\ell=2$ modes.

For a given pair the $\ell=2$ multiplet is offset by the small frequency separation $\dotwo$ from the $\ell=0$ modes, so the $\ell=2$ frequencies are defined by $\nu_{n_\mathrm{p}-1,2} =\nu_{n_\mathrm{p},0}-\dotwo$. The final part in defining the location of the modes in the model is to parameterize the $\ell=0$ mode frequencies $\nu_{n_\mathrm{p}, 0}$, where we use the same asymptotic relation for \pmodes \citep[e.g.,][]{Mosser2015} as in the initial release of \pbjam, which are given as
\begin{equation}
\nu_{n_\mathrm{p}, 0} =\dnu\left(n_\mathrm{p}+\eps_\mathrm{p}+\frac{\alpha_\mathrm{p}}{2}\left(n_\mathrm{p}-n_{\mathrm{max}}\right)^2\right).
\label{eq:asymptotic_p_l0}
\end{equation}
Here $\eps_{\mathrm{p}}$ is a frequency offset in units of $\dnu$, $\alpha_\mathrm{p}$ determines the scale of the second order variation of the mode frequencies with radial order, and $n_{\mathrm{max}}$ is the notional value of a radial order corresponding to $\numax$ defined by $n_{\mathrm{max}}=\numax/\dnu - \eps_{\mathrm{p}}$. This model produces a set of $N_\mathrm{p}$ $\ell=2,0$ mode pairs, which are approximately spaced by $\dnu$ and centered on $\numax$. As in \citet{Nielsen2021}, $N_\mathrm{p}$ is an integer value which must be manually chosen and is usually between 5 and 10 for low and high \snr cases respectively. Here and in the following we use subscript $\mathrm{p}$ to distinguish these parameters as pertaining only to the \pmodes.

The heights of the Lorentzian profiles are defined as 
\begin{equation}
    H_{n_\mathrm{p} \ell m} = \mathcal{E}_{\ell m}\left(i_{\star}\right) V_{0,\ell}^2 H_{\mathrm{E}}\exp\left(\frac{-\left(\nu_{n_\mathrm{p},\ell}-\numax\right)^2}{ 2\,W^2_{\mathrm{E}}}\right).
    \label{eq:modeheights}
\end{equation}
This parametrizes the mode heights as primarily a function of the mode frequency, such that they follow a Gaussian envelope centered on $\numax$, with a height $H_\mathrm{E}$ and a width $W_{\mathrm{E}}$. 

For modes of $\ell >0$ the heights within a rotationally split multiplet are further modulated by $\mathcal{E}_{\ell m}\left(i_{\star}\right)$ which determines the relative visibility of the azimuthal orders \citep[see][for a definition]{Gizon2003}. This visibility factor is unity for modes of $\ell=0$ and for $\ell>0$ ranges between $0$ and $1$ depending on the azimuthal order and the inclination, $i_{\star}$, of the stellar rotation axis relative to the line of sight to the observer. 

The mode heights of a given $\ell$ are further scaled by the ratio, $V_{0,\ell}$, of the spatial response of the mode to that of the $\ell=0$ modes \citep[see, e.g.][]{Bedding1996}. These visibility ratios are kept fixed in the model and we use the values for $V_{0,\ell}$ computed by \citet{Handberg2011} for asteroseismic targets observed by the \kepler mission \citep{Borucki2010}. The mode visibilities depend on the photometric band pass of the observations, but the mode identification in \pbjam only requires approximate values. The \kepler visibilities are therefore still applicable to other observations such as those from the \tess \citep{Ricker2015} mission. 

Lastly, the final part of Eq. \ref{eq:asymptotic_p_l0}, is to parameterize the mode widths. We find that when considering the central few radial orders of the envelope ($N_{\mathrm{p}}\approx10$) the widths of all the modes can to first order be well approximated by a single common width, $\Gamma$ \citep[see, e.g.,][]{Lund2017}. This is sufficient to establish the mode identification for the $\ell=2,0$ modes.

\subsection{Identification of $\ell=1$ modes}
\label{sec:l1modeID}
The addition of the $\ell=1$ modes requires a more flexible description of the mode frequencies. In the following we describe the methods for deriving the $\ell=1$ mode frequencies, whether strongly mixed or not. We label these methods broadly in relation to the evolutionary stage of the stars for which they are intended to be used, namely \MS, \SG and \RG stars. However, these model labels are only indicative of which type of star they are most appropriate for and are not directly related to the structure of the star in question. The \SG model for example is in principle applicable to all types of stars, but is not the most computationally efficient model to use for \MS or \RG stars. Automatically distinguishing each stage based on the observational inputs is often ambiguous at the boundaries between each stage. We therefore leave it to the user to decide when using each method is appropriate. 


Similar to the model for the $\ell=2,0$ modes, the spectrum model for the $\ell=1$ modes is given by a sum of Lorentzian profiles
\begin{equation}
M_{\ell=1}\left(\nu\right) = \sum\limits_{n=1}^{N_\mathrm{p}+N_\mathrm{g}} \sum\limits_{m=-\ell}^{\ell}\frac{H_{n\ell m}}{1+\frac{4}{\Gamma_{n\ell}^2}(\nu-\nu_{n\ell m})^2}.
\label{eq:l1mod}
\end{equation}
Here the number of multiplets to consider is the sum of $N_\mathrm{p}$ \pmodes and $N_\mathrm{g}$ \gmodes,  where $N_\mathrm{p}$ is the same number of \pmodes included in the $\ell=2,0$ model above. The number of \gmodes is given by those that we estimate are likely to couple to these \pmodes based on the prior sample (see Section \ref{sec:prior}). For \MS stars $N_\mathrm{g}=0$, but for \SG stars this increases to $\sim 1-10$, and to potentially several dozens for \RG stars \citep[see, e.g,][]{Mosser2011}. 

The distinction between these cases lies in the treatment of the mode frequencies $\nu_{n\ell m}$, which are described below. However, we use the same parameterization of the mode heights as in Eq. \ref{eq:modeheights}, such that the $\ell=1$ mode heights also follow the Gaussian envelope in frequency. In this step of the algorithm the envelope parameters $H_{\mathrm{E}}$ and $W_{\mathrm{E}}$ are fixed to the median of the posterior distributions from the $\ell=2,0$ model stage. This is an approximation to reduce the dimensionality of the model and again reduce the computational cost of the sampling.

The mode widths for the $\ell=1$ modes are similarly the common width, $\Gamma$, established by the $\ell=2,0$ model. However we scale this by the degree of mixing, $\zeta_{n, l=1}$, (defined below) such that $\Gamma_{n \ell=1}=(1-\zeta_{n \ell=1})\,\Gamma$. This accounts for the coupling to the \gmodes which increases the mode inertia and by extension lengthens the damping timescales compared to the pure \pmodes, under the standard assumption that radiative and convective damping primarily occur in the p-mode cavity \citep{benomar_damping_2014}. In frequency-space mixed modes can therefore have significantly smaller widths, to the point where they may sometimes not be resolvable in short time series such as those from \tess.

\subsubsection{Asymptotic p-mode frequencies}
\begin{figure}
    \centering
    \includegraphics[width=1\linewidth]{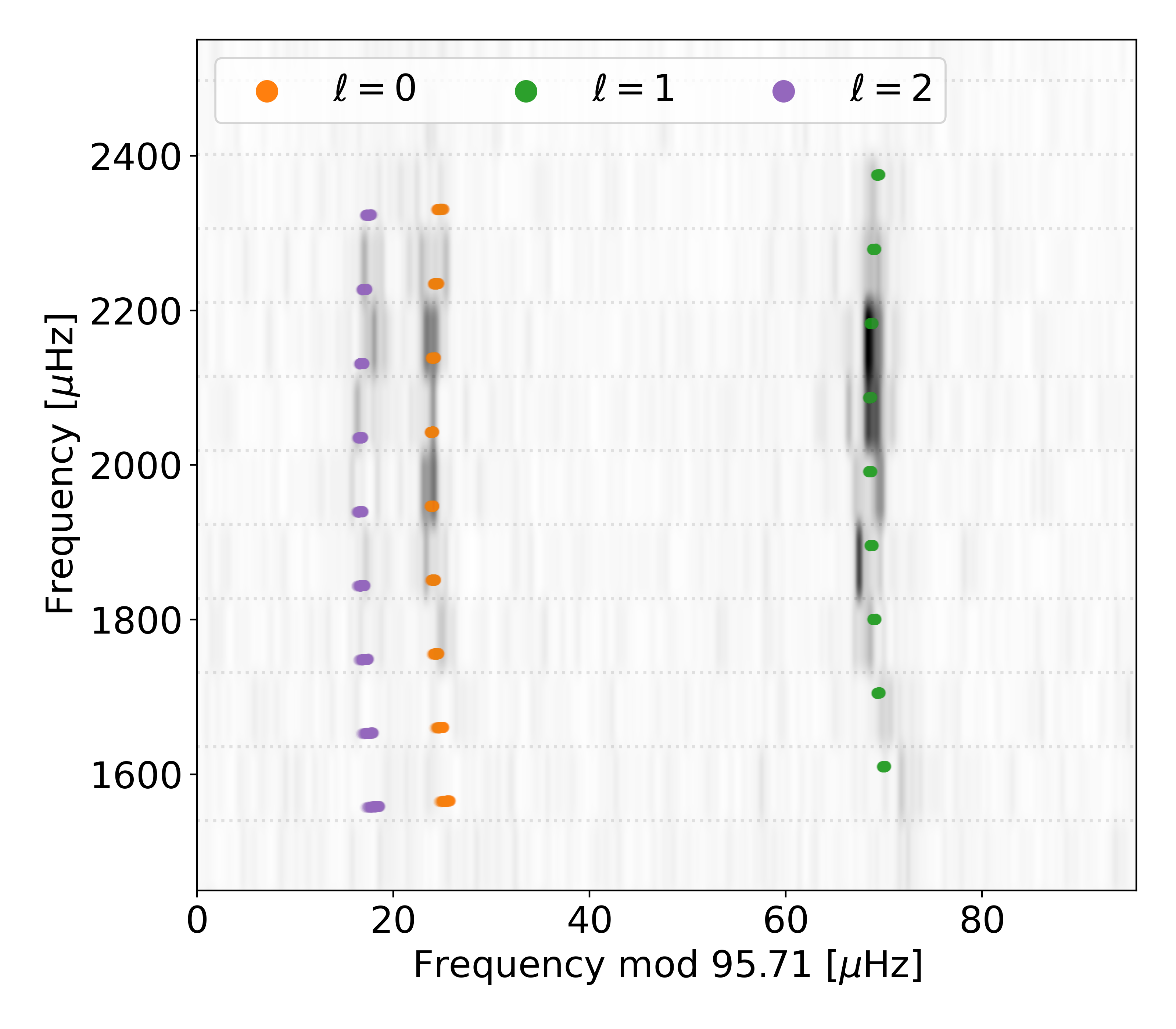}
    \caption{Èchelle diagram of an \MS star KIC~5184732 shown in gray, and the result of applying the \MS model for mode identification. The colored points are samples drawn from the marginalized mode frequency posterior distributions, and so a wider collection of points denotes a larger uncertainty on the location of the mode. For clarity, only modes of azimuthal order $m=0$ are shown. These frequency samples correspond to those used in the top frame of Fig.~\ref{fig:examplespectra} and to construct the models shown in Fig.~\ref{fig:modelSteps}.}
    \label{fig:MSmodeID}
\end{figure}

To parameterize the mode frequencies and identify the likely location of the $\ell=1$ modes we have established three different algorithms which are suitable for either \MS, \SG, or \RG stars. The simplest case is \MS stars where there is little to no coupling to any \gmodes, and so the $\ell=1$ modes follow the asymptotic relation for \pmodes like the $\ell=2,0$ modes. In this case the $\ell=1$ mode frequencies are given by
\begin{equation}
    \nu_{n_{\mathrm{p}},\ell=1} = \nu_{n_\mathrm{p}, \ell = 0} + d_{01},
    \label{eq:asymptotic_p_l1}
\end{equation}
where the $\nu_{n_\mathrm{p}, \ell = 0}$ is established by the $\ell=2,0$ model, and $d_{01}$ is the mean separation between the $(n_\mathrm{p},\ell=0)$ modes and their adjacent $(n_\mathrm{p}, \ell=1)$ modes immediately above in frequency. For notionally pure p-modes, this separation is approximately constant in frequency, and so we leave it as a free variable in this model. Figure~\ref{fig:MSmodeID} shows the result of the mode identification for the \MS star KIC~5184732, where the near constant resulting mode frequencies modulo $\dnu$ appear along almost vertical ridges. Equation \ref{eq:asymptotic_p_l1} also forms the basis for computing the mixed modes' frequencies and so we let $d_{01}$ be a random variable in the following methods as well.

\subsubsection{Few g-modes coupling to many p-modes}
\begin{figure}
    \centering
    \includegraphics[width=1\linewidth]{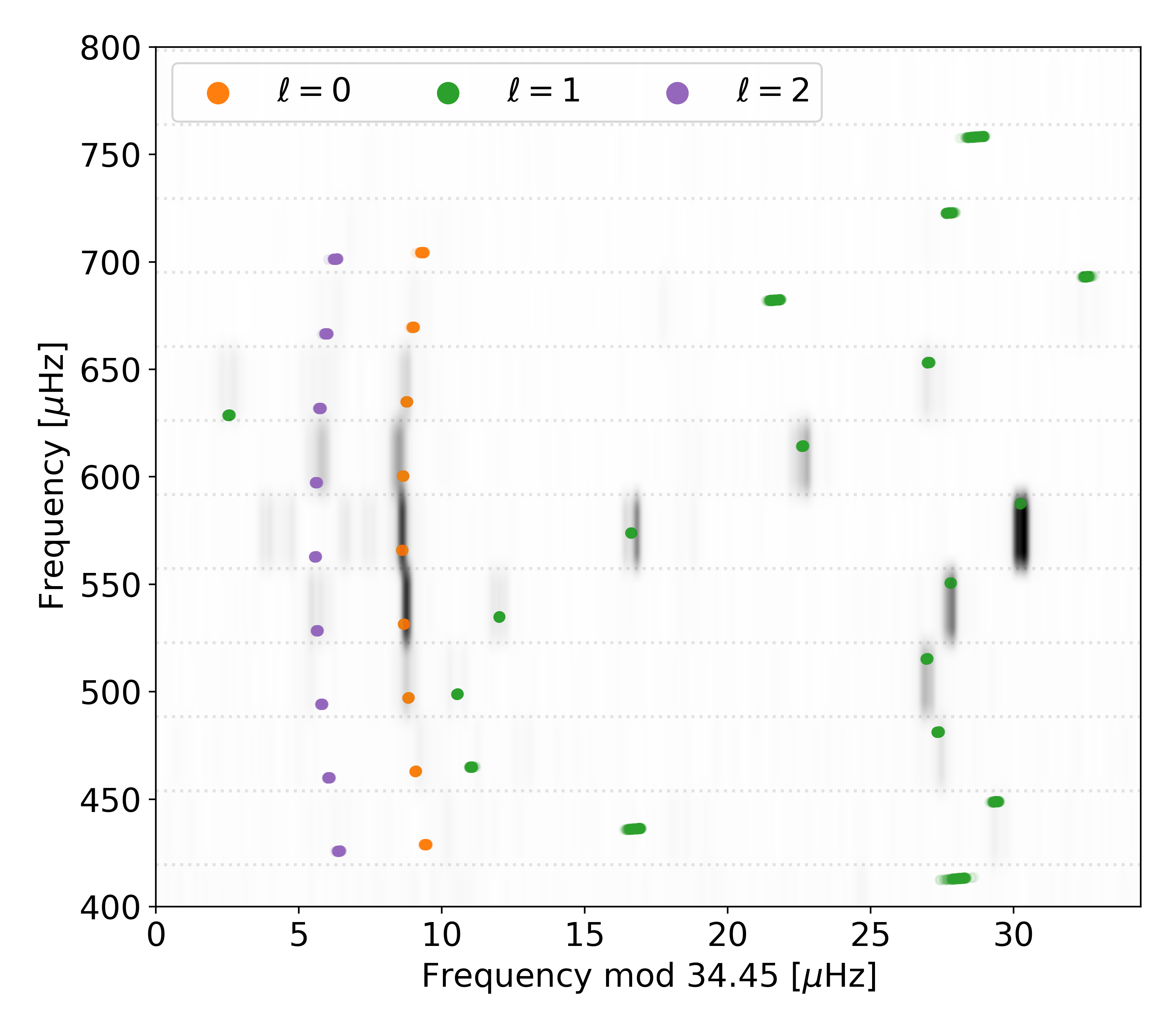}
    \caption{Èchelle diagram of the subgiant star KIC~5723165 (gray) along with the application of the \SG model for mode identification shown in colored points as in Fig.~\ref{fig:MSmodeID}.}
    \label{fig:SGmodeID}
\end{figure}
For \SG stars the \pmode frequencies decrease to the extent that they start to become comparable with the \gmode frequencies. In the following we assume that overtones of pure \gmodes are uniformly spaced in period \citep{Tassoul1980}. In particular, the dipole \gmodes are spaced by $\DPione$, and we parameterize their frequencies as
\begin{equation}
    \nu_{n_{\mathrm{g}},\ell=1} = \left[\DPione\left(n_\mathrm{g} + \eps_\mathrm{g}\right)\right]^{-1}. 
    \label{eq:asymptotic_g_l1}
\end{equation}

However, the non-radial modes of sub- and red giants propagate in two coupled mode cavities, rather than the single mode cavity described by Eqs.~\ref{eq:asymptotic_p_l1} and \ref{eq:asymptotic_g_l1}. Therefore, they are of mixed p-like and g-like character, and do not admit an accurate description with either relation.
 
\cite{Ong2020} identified four regimes of mixed-mode coupling, of which two are specifically relevant to the solar-like oscillators under consideration. Subgiant stars recently coming off the main sequence exhibit only a few g-modes, at low radial order $n_\mathrm{g}$, which couple to many observable p-modes at high $n_\mathrm{p}$. The number of g-modes in the frequency range excited by convective turbulence increases with evolution (so that the observed modes are the result of coupling many g-modes to many p-modes), as do their radial orders, until eventually the mixed-mode spectrum is dominated by g-modes, all of very high radial order.
 
To obtain a model for the mixed mode frequencies we make use of a non-asymptotic description developed in \citet{Deheuvels2010} and \citet{Ong2020}, where we solve a generalized Hermitian eigenvalue equation of the form
\begin{equation}
    \left(\begin{bmatrix}
    -\mathbf{\Omega}^2_\mathrm{p} & \mathbf{L}\\
    \mathbf{L}^\dagger & -\mathbf{\Omega}^2_\mathrm{g}
    \end{bmatrix} + \boldsymbol{\lambda} \begin{bmatrix}
    \mathbb{I} & \mathbf{D}\\
    \mathbf{D}^\dagger & \mathbb{I}
    \end{bmatrix}\right)\mathbf{v} = 0.
    \label{eq:coupling2}
\end{equation}
Here, $\boldsymbol{\Omega}_\mathrm{p}=2\pi\boldsymbol{\nu}_\mathrm{p}$ and $\boldsymbol{\Omega}_\mathrm{g}=2\pi\boldsymbol{\nu}_\mathrm{g}$ are diagonal matrices containing the angular frequencies of the notional pure p- and g-mode frequencies, the eigenvalues $\lambda$ yield the mixed-mode frequencies $\boldsymbol{\nu}=\sqrt{-\boldsymbol{\lambda}}/2\pi$, $\mathbf{\nu}$ is the corresponding eigenvector whose components specify mixed modes as linear combinations of p- and g-modes, and $\mathbf{L}$ and $\mathbf{D}$ are coupling matrices. Given these eigenvectors, the mixing fraction $\zeta$ of each mode may then be computed as
\begin{equation}
    \zeta = \mathbf{v}^T\begin{bmatrix}0 & 0 \\ 0 & \mathbb{I}\end{bmatrix}\mathbf{v}.
\end{equation}
This quantity is constructed so that (by convention) it tends to 1 for pure g-modes, and to 0 for pure p-modes.

The matrix elements $L_{ij}$ and $D_{ij}$, which describe the strength of the coupling between the $i$\textsuperscript{th} \pmode and $j$\textsuperscript{th} \gmode, have been shown by \citet{Ong2020} to depend on the eigenfunctions, $\bm{\xi}$, of the respective p- and g-modes with approximate expressions of the form
\begin{equation}
    L_{ij} \approx \int \mathrm d m\  N^2 (\bm{\xi}_{i, \mathrm{p}}^* \cdot \mathbf{e}_r) (\bm{\xi}_{j, \mathrm{g}} \cdot \mathbf{e}_r);\ D_{ij} \approx \int \mathrm d m\ \bm{\xi}_{i, \mathrm{p}}^* \cdot \bm{\xi}_{j, \mathrm{g}},\label{eq:mat}
\end{equation}
where $N$ is the Brunt-Väisälä or buoyancy frequency \citep[see, e.g.,][]{Aerts2010}, $\mathbf{e}_r$ is the unit vector in the radial direction, and $m$ is the mass coordinate inside the star.

\begin{figure*}
    \centering
    \annotate{\includegraphics[width=.49\textwidth]{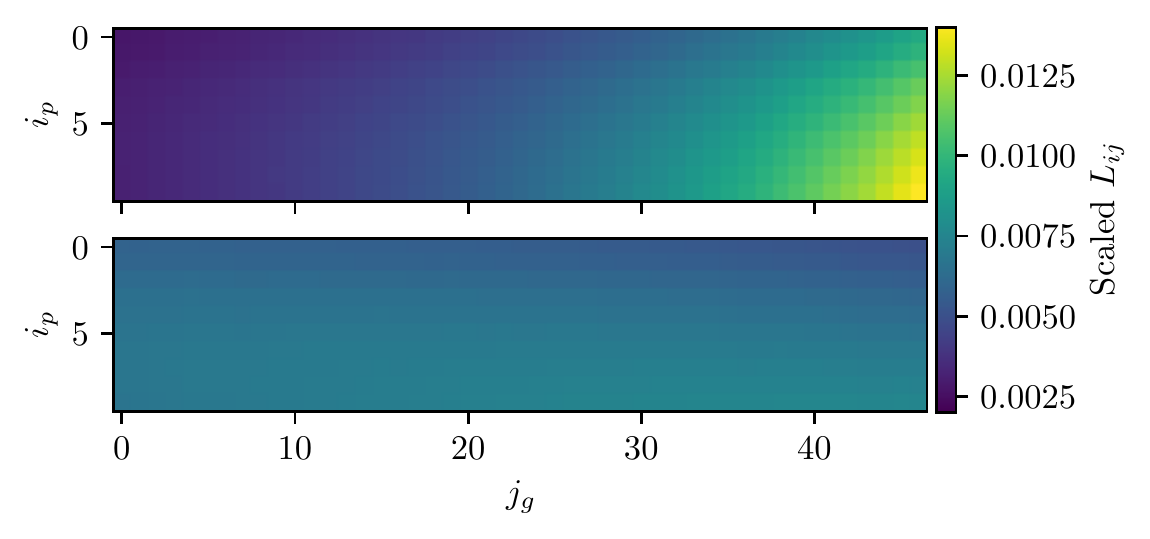}}{\node[white] at (.13, .88){\textbf{(a)}};}
    \annotate{\includegraphics[width=.49\textwidth]{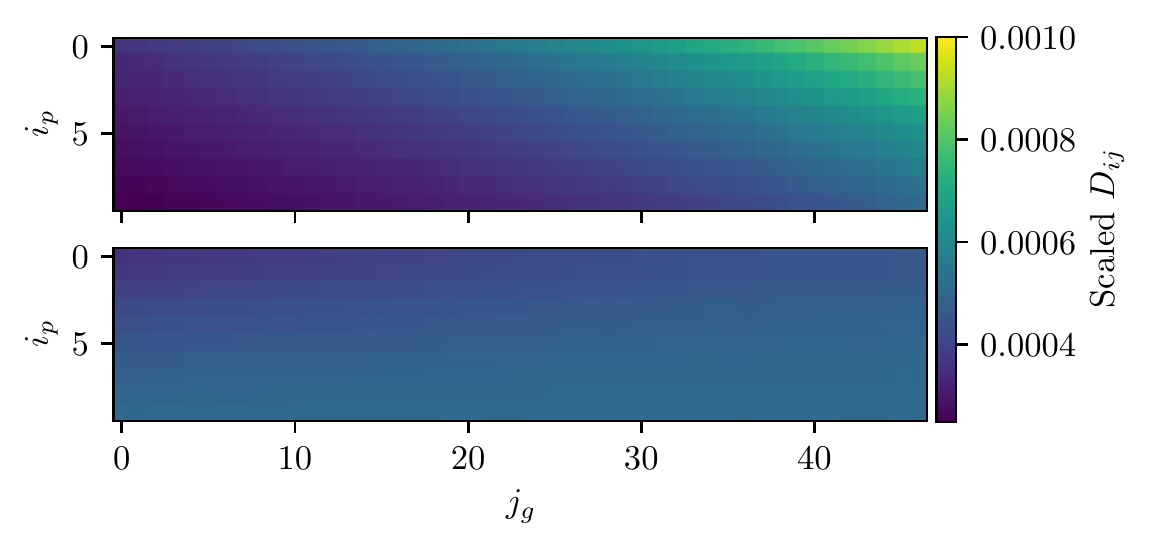}}{\node[white] at (.13, .88){\textbf{(b)}};}
    \caption{Scaled matrix elements $L_{ij}$ and $D_{ij}$ for a red giant stellar structure model (Model I of \citealt{Ong2022}). \textbf{(a)}: Matrix elements $L_{ij}$, scaled to produce dimensionless quantities. The upper panel shows this quantity scaled by $1/\omega_\text{max}^2 = 1/(2\pi\numax)^2$, and exhibits strong dependence on the g-mode frequency index, and weak-dependence on that of the p-modes. The lower panel shows the same matrix elements scaled by $1/\omega_{\mathrm{g}, j}^2$, as described in the main text. The dynamic range of its dependence on the mode indices can be seen to be significantly suppressed. \textbf{(b)}: Similar scalings applied to matrix elements $D_{ij}$, which are already dimensionless. The upper panel shows this quantity exhibiting strong dependence on both p- and g-mode indices. The lower panel shows the same matrix elements scaled by $\omega_{\mathrm{p}, i}/\omega_{\mathrm{g}, j}$, as described in the main text, again showing far milder dependencies on either index.}
    \label{fig:coupling}
\end{figure*}

These matrix elements exhibit strong dependence on the indices (and therefore frequencies) of the associated p- and g-modes, rendering them unsuitable for direct use in any sparse parameterization of this coupling. We illustrate this in \cref{fig:coupling}, where we show the matrix elements for an \RG stellar structure model (Model I of \citealt{Ong2022}). 
Parameterizing the coupling matrices in terms of each individual matrix element directly would require additional free parameters to describe this frequency dependence. We can instead use judicious combinations of these matrix elements with the mode frequencies as our dynamical variables, chosen so as to encode their first-order dependence on frequency. For $L_{ij}$, we note that \cref{eq:mat} indicates that it scales with the square of the Brunt-Väisälä frequency. However, for any given g-mode of fixed $n_\mathrm{g}$, the asymptotic relation for g-mode frequencies indicates that $\omega_\mathrm{g} \propto N$, and so roughly speaking we must have that $L_{ij} \propto \omega_{\mathrm{g}, j}^2$, such that the quantity $L_{ij} / \omega_{\mathrm{g}, j}^2$ is expected to have a significantly reduced dynamic range as shown in  \cref{fig:coupling}. Thus, we will represent the coupling matrix $L$ using a scalar $p_L$ from which matrix elements $L_{ij}$ are recovered as $L_{ij} \approx \omega_{\mathrm{g},j}^2 \cdot p_{\mathrm{L}}$. Likewise, we see that the quantity $D_{ij} \cdot {\omega_{\mathrm{p},i} / \omega_{\mathrm{g},j}}$ has a similarly suppressed dynamic range with frequency. Thus, we represent the coupling matrix $D$ using a scalar $p_{\mathrm{D}}$ from which matrix elements $D_{ij}$ are recovered as $D_{ij} \approx \omega_{\mathrm{g},j}/\omega_{\mathrm{p}_i} \cdot p_{\mathrm{D}}$. By drawing samples of the parameters for the asymptotic p- and \gmode models, together with the pair of mode coupling coefficients $p_{\mathrm{L}}$ and $p_{\mathrm{D}}$ we can construct an approximation of the two coupling matrices in Eq.~\ref{eq:coupling2}. We then solve the eigenvalue problem to recover the mixed mode frequencies $\boldsymbol{\nu}_{n,\ell}$. 

However, this approach only obtains approximate values of the mode frequencies and as Fig.~\ref{fig:coupling} shows, some residual dependence on the p- and g-mode indices remains. This can lead to small discrepancies on the order of $\sim1\mu$Hz between the model and the observed frequencies. To account for this we introduce a Gaussian random variable $\sigma_{n, \ell}\sim \mathcal{N}(0,\,3\%\dnu)$ for each of the $\ell=1$ modes included in the model. The mixed mode frequencies in the spectrum model are therefore given by $\nu_{n, \ell}=\nu^{\prime}_{n, \ell}+\sigma_{n, \ell}$ where the additional term allows for some error in the model on the order of $3\%$ of $\dnu$. The result of applying this method to the \SG star KIC~5723165 is shown in Fig.~\ref{fig:SGmodeID}, where the regular pattern of the $\ell=1$ modes has been disrupted. A pair of mixed $\ell=2$ modes is also visible around the pure $\ell=2$ ridge at $\approx575\muHz$. The methodology applied to treating the coupling of the $\ell=1$ modes can also be extended to encompass $\ell=2$ modes, but this is out of scope for this update and is not currently implemented in \pbjam.
 
In principle, this construction is usable to model the dipole modes in \MS stars, \SG stars with strong coupling, and \RG stars with weak coupling. However, the computational expense of solving Eq. \ref{eq:coupling2} scales by a power of two to three with the number of modes included in the model, and so becomes intractable for \RG stars with dozens of \gmodes per radial \pmode order, while for \MS stars it is an unnecessary addition to the model complexity since no modes are expected to be of mixed character. We therefore only apply this method to \SG stars, and even so limit the number of \gmodes being considered to those that are near the oscillation envelope (see Section \ref{sec:prior} for a description of how this is calculated). 

\subsubsection{Many g-modes coupling to few p-modes}
\label{sec:rgmodel}

\begin{figure}
    \centering
    \includegraphics[width=1\linewidth]{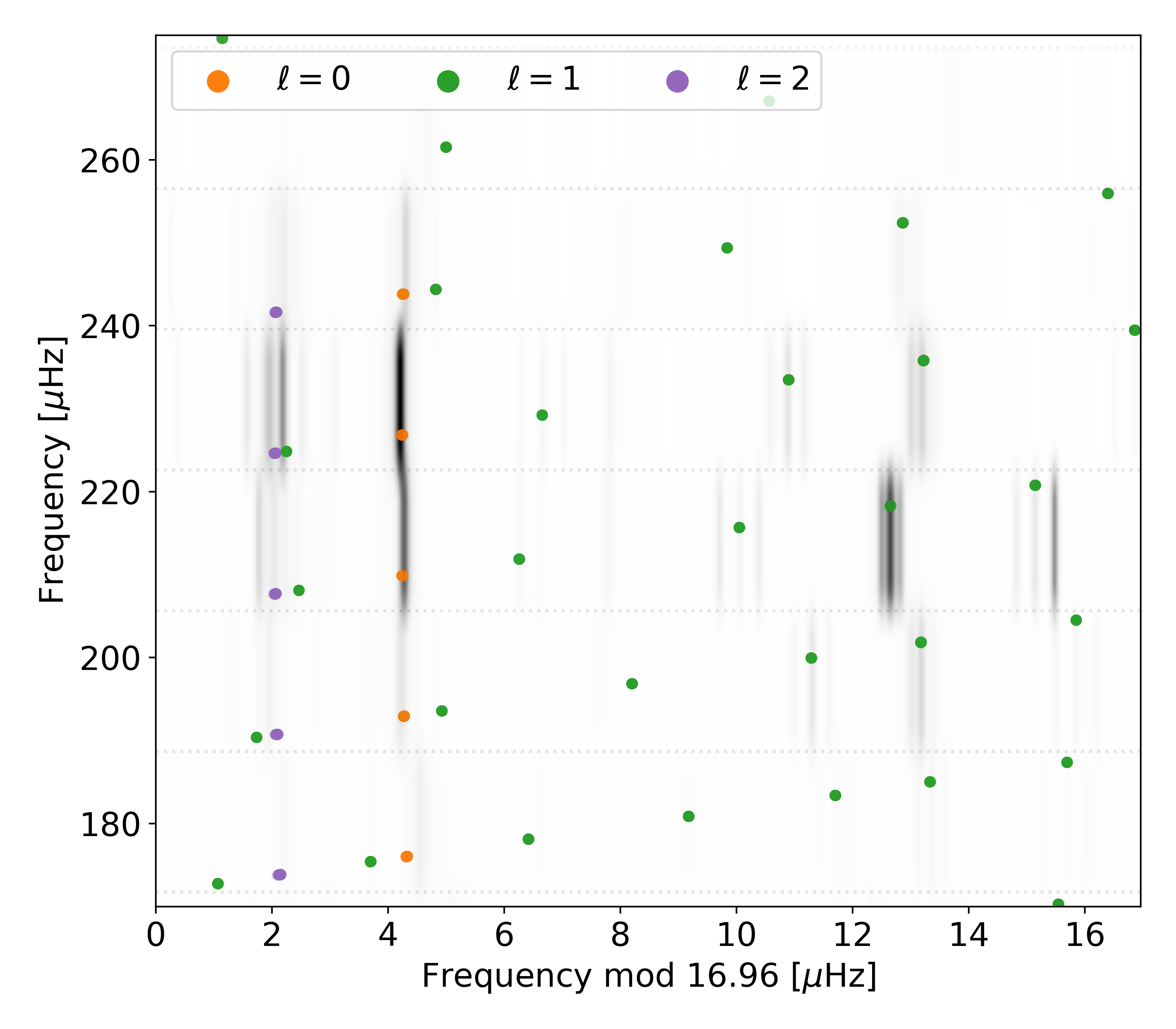}
    \caption{Èchelle diagram of the red giant KIC~4448777 (gray) along with the mode identification derived from the \RG model shown in colored points as in Fig.~\ref{fig:MSmodeID}.}
    \label{fig:RGmodeID}
\end{figure}
For red giants of intermediate advancement up the red giant branch ($\numax\sim100\muHz$), both the p- and g-mode cavities are separately well-described with the JWKB approximation \citep[e.g.][]{Unno1989}, i.e., the typical observable $n_\mathrm{p}$ and $n_\mathrm{g}$ are both $\gg1$. In this case, the mixed-mode frequencies are described by the roots of characteristic equations in the generic form \citep[e.g.][]{shibahashi_modal_1979}
\begin{equation}
    F(\nu) = \tan \theta_\mathrm{p}(\nu) \tan \theta_\mathrm{g}(\nu)- q(\nu) = 0,
    \label{eq:coupling1}
\end{equation}
where $q$ is the coupling strength between modes in the two cavities. We define the phase angles $\theta_\mathrm{p}$ and $\theta_\mathrm{g}$ as
\begin{equation}
    \begin{aligned}
        \theta_\mathrm{p} &\equiv \pi\left({\nu \over \Delta\nu} - \epsilon_{\mathrm{p},\ell=1}(\nu)\right),\\
        \theta_\mathrm{g} &\equiv -\pi\left({1 \over \nu \Delta\Pi_1} - \epsilon_{\mathrm{g},\ell=1}(\nu)\right),
    \end{aligned}
\end{equation}
so that they yield integer multiples of $\pi$ when evaluated at the pure p- or g-mode frequencies specified by \cref{eq:asymptotic_p_l1} and \cref{eq:asymptotic_g_l1}, as required, and moreover both increase monotonically with frequency. For the purposes of numerical sampling, $q$ is treated as a constant in frequency, sampled from a uniform prior $\mathcal{U}(0.01, 0.6)$, along with the other parameters describing the pure g-mode frequencies appearing in \cref{eq:asymptotic_g_l1}.

The mixed-mode frequencies are defined only implicitly through Eq.~\ref{eq:coupling1}, as the roots of the characteristic function $F$, which must in practice be solved for numerically. We implement this root-finding by the damped Halley's method \citep[see, e.g.,][]{Ortega1970}. The pure p- and g-mode frequencies are used as starting guesses for the locations of the roots, and we find that using a damping rate of $0.5$ leads to the root-finding converging in $15$ iterations. This is sufficiently small to allow this method to be used in a sampling algorithm, where the root-finding step is performed to derive the mixed-mode frequencies for each evaluation of the likelihood function. To evaluate the mode widths we again use the common width $\Gamma$ from the $\ell=2,0$ model, but here we compute the mixing fraction $\zeta$ for each mode using the closed-form expression
\begin{equation}
    \zeta(\nu) \equiv \left[1 + {\Delta\Pi_l \over \Delta\nu}{\nu^2 \over q \cos^2 \theta_\mathrm{p}(\nu) + {1\over q}\sin^2 \theta_\mathrm{p}(\nu)}\right]^{-1}.
\end{equation}
Figure \ref{fig:RGmodeID} shows the result of using this approach for mode identification for the \RG star KIC~4448777. 

\subsection{Rotation and inclination}
\label{sec:rotation}
The width of the multiplet is determined by the effective rotation rate sensed by each mode in the multiplet. Modes of p-like character are predominantly affected by the surface rotation \citep[see, for example,][]{Lund2014, Davies2015}, while those of g-like character are more sensitive to the rotation of the deep interior of the star \citep{Deheuvels2012, Deheuvels2014}. Therefore, any resulting mixed modes are split according to the degree of mixing each experiences, and where in the star they are most sensitive to rotation. 

Since \pmodes in \MS stars do not couple with \gmodes, the observed modes almost exclusively experience the rotation of the outer envelope which tends to rotate on the order of days to weeks \citep[see, e.g.,][]{Benomar2015, Santos2021}. The effect is a symmetric splitting of the azimuthal orders $|m|>0$ around the $m=0$ mode in a given multiplet, where all multiplets are assumed to experience the same splitting regardless of radial order or angular degree. We therefore adopt the same formalism for the rotation rate as in the $\ell=2,0$ model, such that $\nu_{n\ell m}=\nu_{n\ell} + m\delta\nu_{\mathrm{env}}$.

In \SG stars where the \pmodes may couple strongly to \gmodes, the resulting splitting of the multiplets must be assumed to be a combination of the rotation rates in their respective oscillation cavities which may vary significantly from one mode to the next. Since our objective in this step of \pbjam is only to provide a mode identification we use a simple two-zone core-envelope model of differential rotation \citep[as in e.g.][]{Goupil2013}. Here the pure \pmodes are split by the envelope rotation parameter $\delta\nu_\text{env}$, and the pure \gmodes by the core rotation parameter $\delta\nu_\text{core}$. The degree of splitting due to the rotation in each zone is a function of the degree of mixing, $\zeta_{n,\ell}$. The splitting in a given multiplet is then assumed to be a linear combination of the form
\begin{equation}
    \nu_{n\ell m} = \nu_{n\ell} + m\left[\zeta_{n\ell}\delta\nu_\text{core} + (1-\zeta_{n\ell})\delta\nu_\text{env}\right].
\end{equation}
While still small compared to for example $\dnu$ and $\dotwo$, the splitting experienced by g-dominated modes can be large compared to the splitting of the p-dominated modes, which experience slower rotation in the envelope compared to the core \cite[see, e.g.,][]{Beck2012}. 

For \RG stars, \pbjam provides the option of using the same linear combination of the core and envelope splittings. However, as shown by \citet{Li2024} this may in some cases lead to a bias in the measured $m=0$ mode frequencies on the order of $0.05-0.1\muHz$. We therefore also include the option to first split the pure p- and g-modes according to the envelope and core rotation rates respectively, such that the pure p- and g-modes are then given by $\nu_{\mathrm{p},i} \mapsto \nu_{\mathrm{p},i} + m \delta\nu_\text{env}$, and the pure g-modes as $\nu_{\mathrm{g},i} \mapsto \nu_{\mathrm{g},i} + m \delta\nu_\text{core}$ \citep{Ong2022}. This is then followed by coupling the rotating p-modes to rotating g-modes via the root-finding algorithm presented in Section \ref{sec:rgmodel}, independently for modes of each azimuthal order. This accounts for the potentially different amounts of mixing experienced by modes of different $m$ over the frequency range spanned by a $\ell=1$ multiplet, which can be a significant fraction of $\dnu$ due to the rapid core rotation of \RG stars \citep{Gehan2018}. This induced asymmetry is estimated from the relative frequencies of the azimuthal orders in the mode identification stage, and are passed forward as fixed quantities to the detailed peakbagging stage (described in Section~\ref{sec:peakbag}). 

Since this approach performs the mode coupling calculations several times, it is more computationally expensive than the linear approach which only does so once. However, by fully accounting for near-degeneracy effects, it has the ability to produce asymmetrically split multiplets \citep[][]{Deheuvels2017,Ong2022}, which are required to interpret the two rotation parameters as being estimates of the physical rotation rate, particularly with increasing evolution up the red giant branch. This effect diminishes for less evolved stars, so while modes in \SG stars can in principle experience a small difference in coupling of the azimuthal orders, the resulting asymmetry is likely not large enough to cause a significant bias, and so the \SG model does not account for this potential asymmetry. 

\subsection{Bayesian inference of the mode identification}
For the purposes of mode identification we seek to sample the posterior probability distribution given by
\begin{equation}
\ln{P(\boldsymbol{\theta}|D)}\propto \ln{\mathcal{L}\left(D|\boldsymbol{\theta}\right)} + \ln{P(\boldsymbol{\theta})},
\label{eq:post}
\end{equation}
where $\ln{\mathcal{L}\left(D|\boldsymbol{\theta}\right)}$ is the likelihood which provides the probability of observing the data given a sample of the model parameters $\boldsymbol{\theta}$, and $\ln{P\left(\boldsymbol{\theta}\right)}$ is the prior probability of drawing that parameter sample. 

As in \citet{Nielsen2021} we include an additional set of observational parameters $O=\{\Teff, \bprp\}$. While these parameters do not directly appear in Eqs.~\ref{eq:l20mod} or \ref{eq:l1mod}, they are strongly correlated with, for example, $\epsilon_\mathrm{p}$ or $\Delta\Pi_1$. These additional observational parameters therefore help inform the model parameters which are otherwise not well constrained by the \PDS itself. The likelihood function is then given as
\begin{equation}
    \ln{\mathcal{L}\left(D|\boldsymbol{\theta}\right)} = \ln{\mathcal{L}\left(S|\boldsymbol{\theta}\right)} + \ln{\mathcal{L}\left(O|\boldsymbol{\theta}\right)},
\end{equation}
where $\ln{\mathcal{L}\left(O|\boldsymbol{\theta}\right)}$ is a sum of normal distributions with a mean of the observed quantity and a standard deviation corresponding to the observational uncertainty. 

In the first stage of the mode identification process, where we sample the parameters of the $\ell=2,0$ model, we assume that the power density is Gamma-distributed with a mean and standard deviation of a limit spectrum that has the functional form equivalent to $M_{\ell=2,0}$ (Eq.~\ref{eq:l20mod}). The joint likelihood of observing the power density across a set of frequency bins given the model is then \citep[e.g.,][]{Anderson1990}
\begin{equation}
 \ln{\mathcal{L}\left(S|\boldsymbol{\theta} \right)} = - \sum\limits_{j=1}^J {\left(\ln M\left( {\boldsymbol{\theta} ,\nu _j } \right) + \frac{{S_j }}{{M\left( {\boldsymbol{\theta} ,\nu _j } \right)}}\right)},
 \label{eq:spectrumlnlike}
\end{equation}
where $S_j$ is the power density at a frequency $\nu_j$. 

In the second stage of the mode identification process, where we identify the $\ell=1$ modes, we divide the \PDS by a mean $\langle M_{\ell=2,0} \rangle$ of $30$ draws of the resulting model posterior distribution. In Eq.~\ref{eq:spectrumlnlike} $S_j \mapsto S_j/\langle M_{\ell=2,0, j} \rangle$, which is also Gamma-distributed but where the limit spectrum is now assumed to be given only by $M_{\ell=1}$ (Eq.~\ref{eq:l1mod}). 

\subsubsection{The priors}
\label{sec:prior}
\begin{figure}
    \centering
    \includegraphics[width=\linewidth]{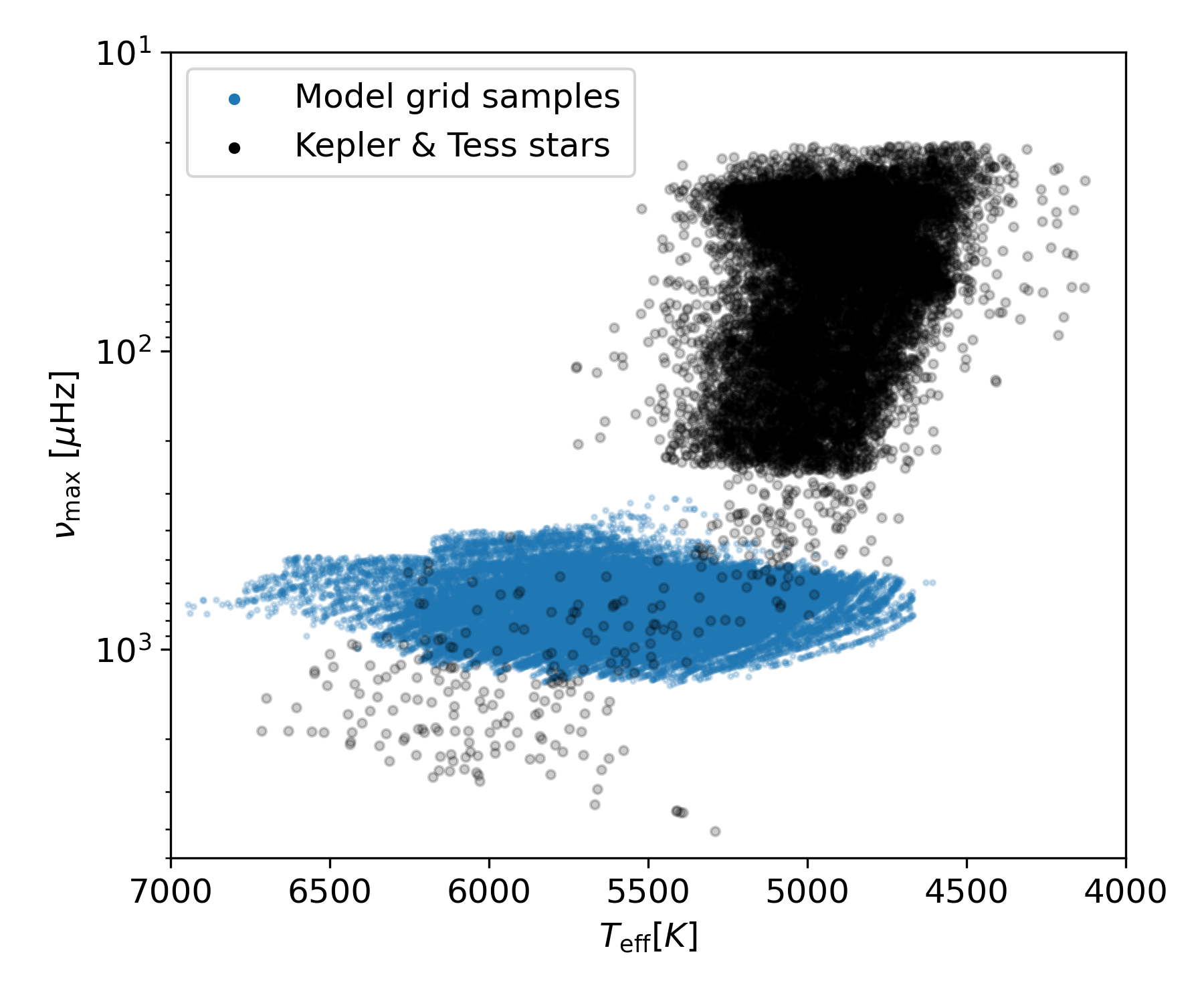}
    \caption{The frequency of $\numax$ of the prior sample represented as a function of $\Teff$. The black points are stars in the sample that have been observed with \kepler or \tess, and for which the $\ell=2,0$ model parameters are complete. The blue points are model grid samples taken from \citet{lindsay_fossil_2024} to supplement the $\ell=1$ model parameters for \SG stars.}
    \label{fig:priorSample}
\end{figure}

The remaining term in Eq. \ref{eq:post} is the log-prior probability, $\ln{P(\boldsymbol{\theta})}$. To compute this probability we take a similar approach as in \citet{Nielsen2021}, where a kernel density estimate of a large sample of previous measurements of the model parameters is used to construct a prior probability density. The purpose of this approach is to encode the current state of knowledge about the model parameters in a nonparametric way, and allow them to be updated as new measurements are performed. The sample currently consists of 13,413 \kepler targets, 288 \tess targets. The set of parameters used for the $\ell=2,0$ model is complete for stars from $\numax\sim30\textup{--}4000\muHz$. However, this sample only contains  $\approx40$ sub-giants with resolvable modes \citep[see, e.g.,][]{Li2020, Appourchaux2020}. We have therefore included an additional 30,812 stellar model grid samples from \citep{lindsay_fossil_2024} for $\numax\gtrsim 300\muHz$. These are used to define the prior density for the \gmode parameters for the sub-giant and red giant models in this frequency range. These two component of the prior sample are shown in Fig. \ref{fig:priorSample}.

Given the large number of parameters that are used in the different models, we apply the dimensionality reduction method based on principle-component analysis described in \citet{Nielsen2023}. This approach relies on the majority of parameters in the spectrum model depending on the same smaller set of fundamental stellar properties, such as temperature, mass, and radius \citep[see, e.g.,][for several examples of asteroseismic scaling relations]{Chaplin2014}. This dependency means that the model parameters are all highly correlated. We exploit this by first selecting a subsample of the previously measured model parameters surrounding the target by picking the closest $100$ targets in the prior sample in terms of the input $\numax$, $\dnu$, and $\Teff$. We then compute the eigenvectors of the covariance matrix of this subsample, where the eigenvectors are a linear combination of the model parameters and represent a set of latent parameters which are orthogonal by construction. By projecting the prior sample onto this space we can then compute a set of one-dimensional kernel-density estimates of the sample, which are used as prior probability densities during the parameter sampling. The orthogonality of the latent parameter space makes constructing and evaluating the prior simpler, but has the added benefit that we may identify the linear combinations (eigenvectors) of model parameters which explain the least variance in the full model space. These are then discarded, thus reducing the volume and dimensionality of the relevant parameter space. We then sample this lower-dimensional latent space and project each sample back into the full model space, after which the model is computed, and the posterior probability is evaluated. 

It is only beneficial to apply this method to parameters that are expected to be highly correlated. This excludes the lowest frequency Harvey-like term and the shot noise level in Eq.~\ref{eq:l20mod}. While some fraction of the power at low frequencies can be expected to be due to intrinsic stellar variability from magnetic activity and rotation, a large part of the observed power must be expected to also be due to residual, uncorrected instrumental variability which is not expected to be correlated with the stellar variability. 

Similarly, the rotation and inclination parameters are not expected to strongly correlate with the remaining model parameters, except across a wide range of $\numax$. These parameters are therefore not included when computing the prior probability density in the latent space. We therefore define separate prior distributions for each of the parameters, which are shown in Table~\ref{tab:instrumental_priors}.
\begin{table}[]
    \centering
    \begin{tabular}{cc}
     Parameter & Prior PDF \\
      \hline
       $\log_{10}{a_{\mathrm{I}}}$  & $\mathcal{N}\left(\mu_{\mathrm{I}}, 1  \right)$ \\ 
       $\log_{10}{b_{\mathrm{I}}}$  & $\beta_{a}^{b}\left(-1, 2\right)$ \\ 
       $c_{\mathrm{I}}$ & $\beta_{a}^{b}\left(1.5, 2.5\right)$ \\ 
       $\log_{10}{S}$ & $\mathcal{N}\left(\mu_S, 0.1\right)$ \\ 
       $\log_{10}{\delta\nu_{\mathrm{env}}}$ & $\mathcal{U}(-2, 0.5)$\\
       $\log_{10}{\delta\nu_{\mathrm{core}}}$ & $\mathcal{U}(-2, 0.5)$\\
       $i_{\star}$ & $\sin{i_{\star}} \text{ for $0 \leq i_{\star} \leq \pi/2$}$
    \end{tabular}
    \caption{The prior probability density functions for the parameters which are not included in the dimensionality reduction. The arguments to the functions denote the location and scale parameters of the distributions. The mean values $\mu_I$ and $\mu_S$ are estimated based on the observed spectrum. The $\beta$ distributions are described by shape parameters $a=b=1.2$. The prior on the stellar inclination, $i_{\star}$, is a sine function truncated to only be non-zero between $i_{\star}=0$ and $i_{\star}=\pi/2$.}
    \label{tab:instrumental_priors}
\end{table}
 
With the prior defined as above, we can also estimate which \gmode radial orders to consider in the mode identification (Section \ref{sec:l1modeID}). We first sample the prior distribution subject only to the constraint of the observed $\numax$, $\dnu$ and $\Teff$. This provides marginal probability densities of $\Delta\Pi_1$ and $\epsilon_\mathrm{g}$, where we use the respective median values of these densities to compute an initial guess for the \gmode radial orders and frequencies using Eq. \ref{eq:asymptotic_g_l1}. When computing the mode coupling we only retain the \gmodes that fall within $5\,\dnu$ below and above the lowest and highest $\ell=0$ mode frequencies respectively. This assumes that this set of \gmodes is sufficient to account for the mixing observed in the frequency range around the envelope, and further that they do not produce visible peaks in the \PDS at frequencies beyond the envelope. 

\subsubsection{Sampling the mode identification model parameters}
To evaluate the marginal probability density of each of the model parameters, we draw samples from the prior volume and evaluate the posterior probability density (Eq.~\ref{eq:post}) at each sample. However, since the models used for the mode identification contain several Lorentzian profiles to describe the mode power, the likelihood function (Eq.~\ref{eq:spectrumlnlike}), and thereby the posterior density must be expected to be multimodal as well. For example, varying $\numax$ by integer multiples of $\dnu$ can produce multiple sets of models that have a high likelihood of explaining the data. To account for this we use the \texttt{Dynesty} nested sampling library for Python \citep{Speagle2020} to map the posterior distribution, since nested sampling has been shown to accurately map multimodal posterior distributions \citep[see][]{Skilling2004, Feroz2013}. In the mode identification stage we use $50$ live points per parameter in the model being sampled, which we find returns consistent results between several random initializations of the sampler with the same model and data set. \texttt{Dynesty} estimates the model evidence at each sampled step in parameter space, and as more samples are accumulated, the change in evidence decreases. We use the default settings in \texttt{Dynesty} for the change in log-evidence in order to terminate the sampling, which occurs when the change in evidence falls below the percent level. 

\section{Detailed Peakbagging}
\label{sec:peakbag}
\begin{figure*}
    \centering
    \includegraphics[width=0.5\linewidth]{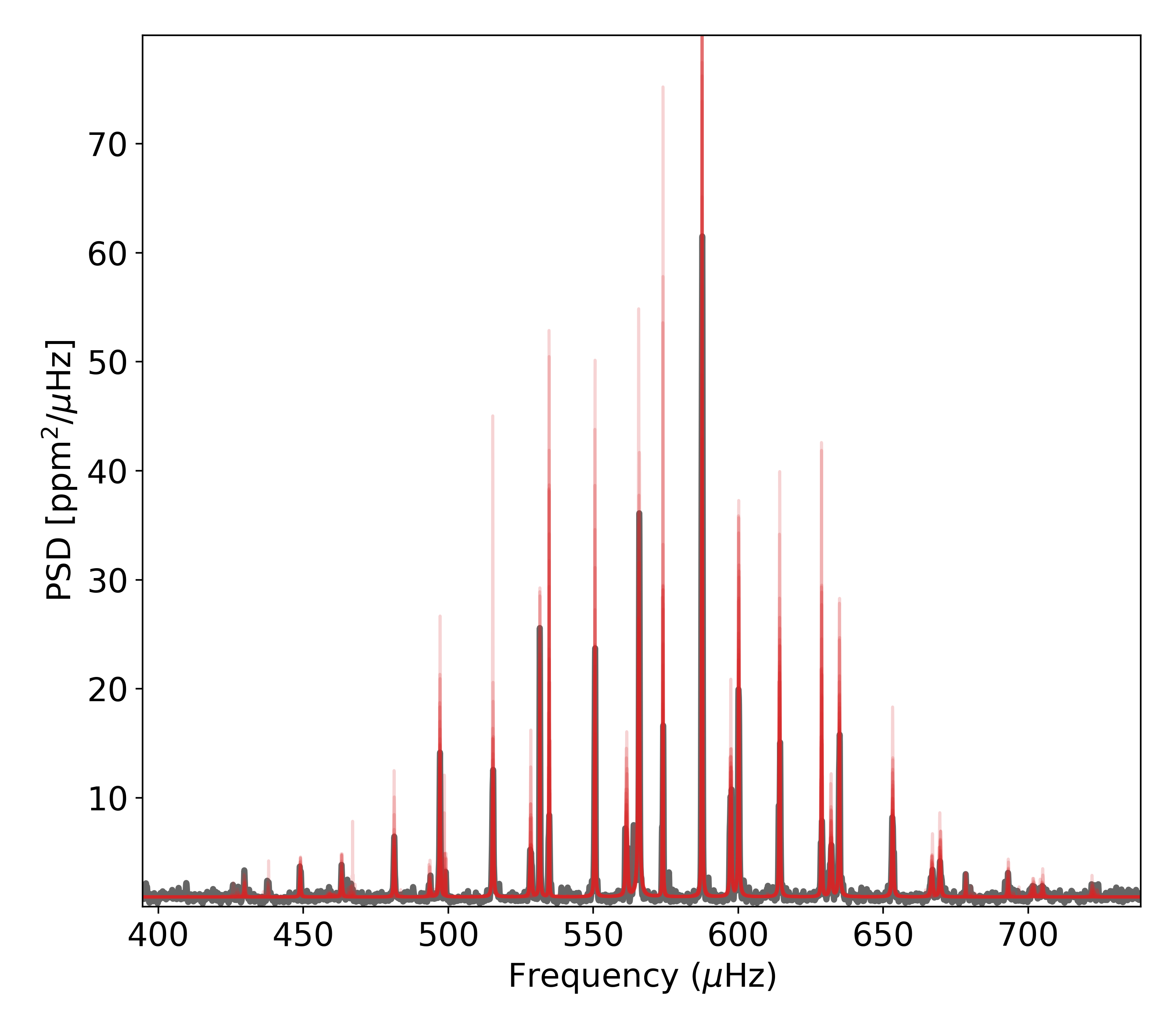}
    \includegraphics[width=0.49\linewidth]{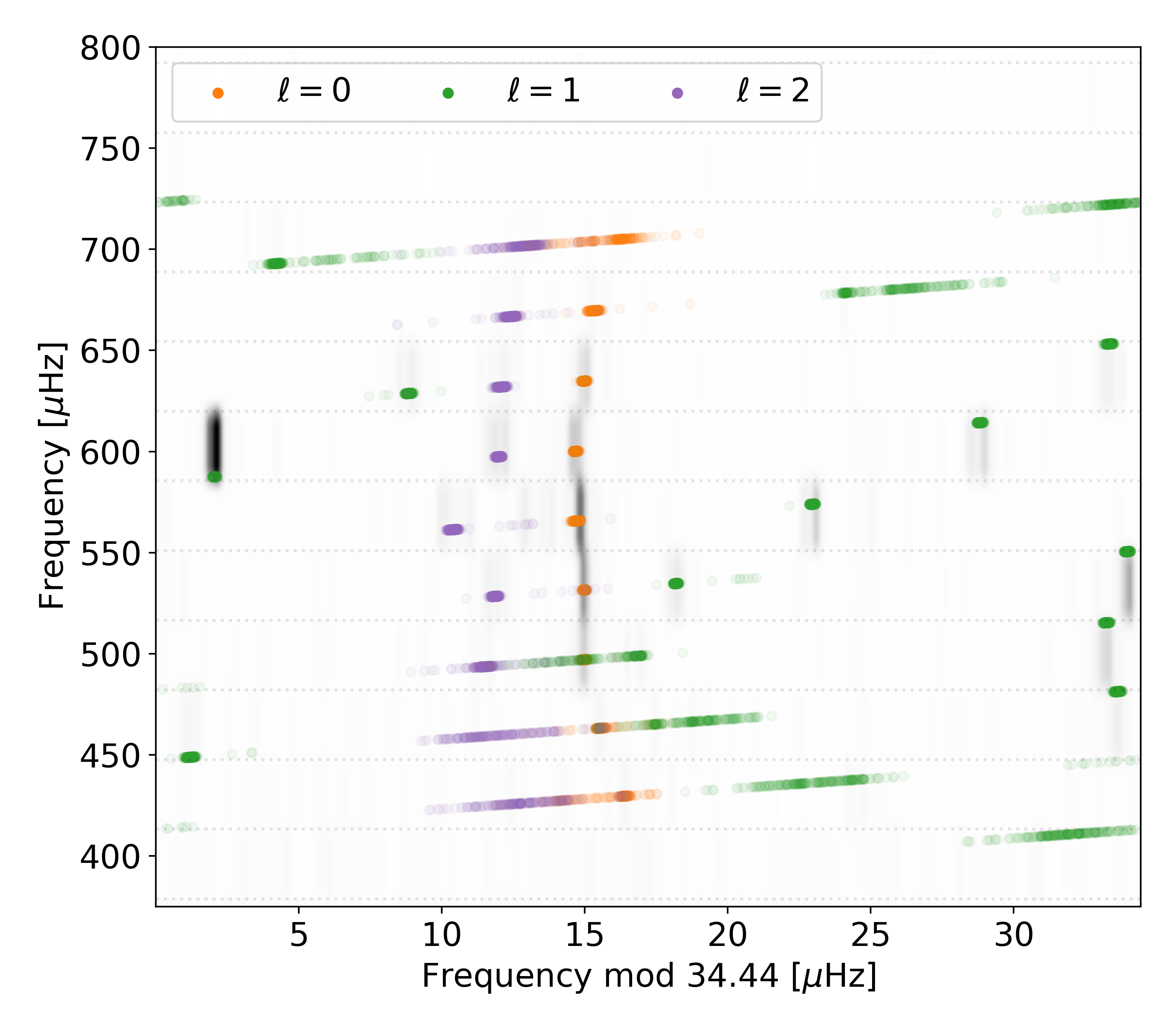}
    \caption{Left: The result of the detailed peakbagging performed for the subgiant KIC~5723165. The smoothed spectrum is shown in black, with models generated from 30 samples of the posterior distribution shown in red. Right: Èchelle diagram like that shown in Fig.~\ref{fig:SGmodeID}, but with the mode frequencies derived from the peakbagging posterior distribution.}
    \label{fig:peakbagging}
\end{figure*}
The second stage of \pbjam, following the mode identification, is the application of a more detailed peakbagging model with fewer constraints. This allows the mode frequencies to be adjusted by up to a few $\muHz$ to accommodate small errors or effects such as acoustic glitches that are not captured by the mode identification stage\footnote{The peakbagging stage of \pbjam allows for the inclusion of additional modes which may not have been included in the mode identification, such as $\ell=3$ modes or modes far from $\numax$, may be manually added to the list of inputs at this stage.}. The mode frequencies, heights, and widths are now left as independent free variables, only being subject to priors that are informed by the mode identification stage. 

\subsection{The peakbagging model}
The spectrum model used in the final peakbagging stage is similar to those constructed during the mode identification, but describes the \snr spectrum instead of the \PDS, and is given by  
\begin{equation}
M\left(\nu\right) = \sum\limits_{n=1}^{N_\mathrm{p}+N_\mathrm{g}} \sum\limits_{\ell=0}^{\ell=2} \sum\limits_{m=-\ell}^{\ell}\frac{s_{n\ell}\mathcal{E}_{\ell m}\left(i_{\star} \right)}{1+\frac{4}{\Gamma_{n,\ell}^2}(\nu-\nu_{n\ell m})^2} + W.
\label{eq:pbmodel}
\end{equation}
However, this spectrum model contains all of the modes from the $\ell=2,0$, and $\ell=1$ models in the mode identification, so that the model accounts for any correlations between modes due to potential proximity in terms of frequency. The height of the Lorentzian profile, which is now defined in terms of the mode \snr, and the widths are left as independent parameters for each multiplet $(n,\ell)$. The mode frequencies are now defined as 
\begin{equation}
\nu_{n\ell m} = \nu_{n\ell} + (m^2 \alpha_{n\ell} + m)\,\left(\zeta_{n\ell}\delta\nu_\text{core} + (1-\zeta_{n\ell})\delta\nu_\text{env}\right),
\end{equation}
such that the model can still account for variation in the splitting experienced by modes with different degrees of mixing. Here $\zeta_{n,\ell}$ is kept fixed during the sampling, and is inherited from the mode identification stage. We also introduce an asymmetry parameter $\alpha_{n \ell}$ to account for the possibility that the splitting of the multiplet is asymmetric in frequency around the $m=0$ mode. This parameter is defined as a fraction of the rotation rate felt by the mode, and ranges between $-1$ and $1$, where $0$ denotes a symmetric multiplet and $-1$ or $1$ means the $m=\pm1$ mode is equivalent to the $m=0$ mode respectively. The asymmetry is derived from the azimuthal orders used in the models in the mode identification stage, and is kept fixed in this stage of \pbjam. This parameterization of the asymmetry only extends to $\ell=1$, and is only non-zero when using the \RG model where the rotation is included using the method by \citet{Li2024} which can account for the differing coupling of the individual azimuthal orders.  

\subsubsection{The peakbagging priors}
The priors on each parameter are now defined as shown in Table \ref{tab:peakbagpriors}. The mode frequency priors are normal distributions centered on the frequencies derived from the mode identification, but with a width of $3\%$ of $\dnu$. This construction of the mode frequency priors does not exclude the possibility that, for example, the $\ell=2$ and $\ell=0$ mode frequencies can swap places. Therefore, as an additional weak prior we apply the constraint that for a list of mode frequencies the differences must always be positive so that they appear with the same ordering as supplied by the mode identification stage. 

The priors on the mode heights are normal distributions in logarithmic \snr, where the mean is estimated by dividing the mode heights from the mode identification stage by the local estimate of the background noise level. The mode width priors are also normal distributions in logarithmic frequency, centered on the estimates from the mode identification. Similar to the mode identification stage, for $\ell=1$ modes of mixed character, the priors on the mode widths are scaled by $\zeta_{n,\ell=1}$ to account for the higher mode inertia. 

\begin{table}[]
    \centering
    \begin{tabular}{cc}
         Parameter & Prior PDF \\
         \hline
         $\nu_{n\ell}$ &  $\mathcal{N}\left(\nu^{\mathrm{\prime}}_{n\ell}, 0.03\dnu^2\right)$ \\
         $\log_{10}{h_{n\ell}}$ & $\log{\mathcal{N}\left(\log{h^{\mathrm{\prime}}_{n\ell}}, 0.5\right)}$ \\
         $\log_{10}{\Gamma_{n\ell}}$ & $\log{\mathcal{N}\left(\log{\Gamma^{\mathrm{\prime}}}, 0.1\right)}$ \\
         $\log_{10}{\delta\nu_{\mathrm{env}}}$ & $\mathcal{U}(-2, 0.5)$\\
         $\log_{10}{\delta\nu_{\mathrm{core}}}$ & $\mathcal{U}(-2, 0.5)$\\
         $i_{\star}$ & $\sin{i_{\star}} \text{ for $0 \leq i_{\star} \leq \pi/2$}$ \\
         $\log_{10}{W}$ & $\mathcal{N}(0, 0.01)$
    \end{tabular}
    \caption{Prior probability densities used in the detailed peakbagging stage of \pbjam. The priors on $\nu_{n\ell}$, $h_{n\ell}$, and $\Gamma_{n\ell}$ are normal distributions centered on the respective results from the mode identification stage. We use the same prior on the stellar inclination, $i_{\star}$, as in the mode identification stage.}
    \label{tab:peakbagpriors}
\end{table}

\subsection{Sampling the peakbagging model parameters}
To sample the parameters in Eq.~\ref{eq:pbmodel} we include two separate options. As in the initial release of \pbjam, we include the option to sample with the \texttt{emcee} library \citep{Foreman-Mackey2013}. The sampling is performed using a Markov Chain Monte-Carlo algorithm. The number of chains is set to six times the number of independent parameters in the model, which for most stars equates to $200-400$ chains. The default choice in \pbjam is to use the differential-evolution move strategy \citep{Nelson2014b} with the affine-invariant ensemble sampler \citep{Goodman2010}. The sampling initially runs for a sufficient number of steps for the chains to reach a stationary state in the vicinity of the main mass of the posterior distribution. A useful indicator of this is when the integrated autocorrelation time, $T_{\mathrm{ACF}}$ reaches an approximately constant value \citep{Sokal1997}. Here we use the mean value, $\langle T_{\mathrm{ACF}} \rangle$ across all the chains, and all parameters in the spectrum model, which is computed periodically during the sampling. When the relative change $\langle T_{\mathrm{ACF}} \rangle$ falls below $0.005\%/\mathrm{step}$, this initial `burn-in' phase is considered complete. This occurs after approximately $10,000-20,000$ steps. At this point the $T_{\mathrm{ACF}}$ is estimated for a final time, which is then used to thin subsequent samples to ensure that they are approximately independent of each other. This is continued until at least $5000$ independent samples are obtained, however we note that more samples are likely required when using models consisting of several tens of modes as the dimensionality and thus volume of the posterior space becomes larger.

The second choice of sampler is the \texttt{Dynesty} nested sampler as is used in the mode identification stage. This sampler is better suited for multimodal posterior distributions than the affine-invariant sampler in the \texttt{emcee} library. While this is likely not a problem for \MS or even \SG stars, the mixed mode density in \RG stars combined with the rotational splitting (see Fig.~\ref{fig:examplespectra}) means that the posterior may be multimodal. This comes at the cost of the sampling efficiency which for nested sampling algorithms decreases strongly with the number of dimensions in the model. The time required to accurately map the posterior distribution should therefore be expected to be substantially longer than with the \texttt{emcee} sampler, when considering several tens of modes. This sampler is therefore likely best suited for \RG stars where the width of the oscillation envelope is only a few tens of $\muHz$, and so the model can be limited to a relatively small range in frequency to speed up the time to evaluate the posterior probability at each sample. As in the mode identification stage, we use the default stopping criterion as defined in the \texttt{Dynesty} documentation, where the relative change in model evidence must reach a predefined value. We also apply the same rule of using $50$ times the number of dimensions in the model, however due to computational constraints we place an upper limit of $3000$ live points which yields on the order of $10^5$ independent samples of the posterior distribution. 

Figure~\ref{fig:peakbagging} shows the result of the detailed peakbagging for the \SG star KIC~5723165. Compared to the èchelle diagram in Fig.~\ref{fig:SGmodeID}, the priors on the mode frequencies are now not as restrictive since we no longer require the modes to adhere to any particular functional form. Therefore, the distribution of samples for each mode frequency is wider, and significantly so further from $\numax$, where the posterior distribution becomes comparable to the $3\%$ of $\dnu$ which we use as the width of the prior. This means that, far from $\numax$, the spectrum does not provide additional information compared to the prior established from the mode identification.

\section{Conclusions}
\label{sec:conclusion}
With this update to the \pbjam open-source software package we introduce tools for automatically identifying modes of angular degree $\ell=1$. This is done using three separate methods: the simple asymptotic relation for \pmodes for \MS stars, a frequency-dependent coupling-matrix formalism for \SG stars, and uniform coupling for \RG stars. This allows \pbjam to perform a more comprehensive identification of modes in the spectrum of a solar-like oscillator. We also add additional model components to account for the background noise level, and the core and envelope rotation rates, which allows \pbjam to be used in a wider context than measuring mode frequencies. 
 
Currently, one of the main limitations of \pbjam is the simplicity of the asymptotic relation used to estimate the location of the modes, which forms the basis for all the analysis of the spectrum. Close to $\numax$ this relation is adequate to let the final peakbagging stage find the peaks and obtain more precise frequency estimates. Further from $\numax$, however, the accuracy of the asymptotic relation is reduced since it does not include higher order terms to account for, e.g., acoustic glitches. The number of radial orders to consider is therefore limited to $N_\mathrm{p}\approx10$. For high \snr targets where more radial orders are visible, modes further from $\numax$ must therefore be manually added to the list which are passed to the final peakbagging stage. Suitably parameterizing a model and defining a prior distribution for an acoustic glitch model \citep[e.g.][]{Lyttle2023} would allow the mode identification to be extended further to lower frequencies, thereby increasing the number of modes that \pbjam can produce frequency estimates for. 

Another avenue for improving \pbjam is through the use of multiple data sets. Currently \pbjam is only capable of treating a power density spectra from a single intensity time series from instruments such as \kepler or \tess. This can be extended by, for example, computing the joint likelihood of observing the same set of frequencies given two or more independent power density spectra. While this does not necessarily improve the ability to resolve oscillation frequencies, it can mitigate the effect of having only one realization of the oscillations and noise from one time series, potentially improving the \snr of the modes. This is particularly applicable to faint targets or those that have only been observed for a short time. In such cases the spectra may not yield enough information to constrain, for example, the location of the less visible $\ell=2$ modes which can be used to constrain stellar age. The \tess mission is continuously producing new sets of photometric time series for stars across the sky, but \citet{Hatt2023} showed that while several hundred \SG stars show solar-like oscillations, it is typically only the brightest \MS ($V\leq 6$, Lund et al. in prep.) stars that are suitable for asteroseismic analysis. Combining \tess datasets with any of those targets observed by, for example, K2 \citep{Howell2014, Lund2024} or ground-based observations in radial velocity \citep[for example,][]{Grundahl2017, Knudstrup2023} could improve the asteroseismic inference. Similarly, for the upcoming PLATO mission \citep{Rauer2024}, \citet{Nascimbeni2022} suggested that a field of view may be chosen which overlaps with that of the nominal \kepler mission. Combining observations from stars observed in both missions could provide a higher yield of asteroseismic targets, particularly toward the cooler \MS which are favorable targets for exoplanet searches. 

\section{acknowledgments}
MBN, EJH, WJC, and GRD acknowledge the support from the UK Space Agency. JMJO acknowledges support from NASA through the NASA Hubble Fellowship grant HST-HF2-51517.001, awarded by the Space Telescope Science Institute (STScI). STScI is operated by the Association of Universities for Research in Astronomy, Incorporated, under NASA contract NAS5-26555. GRD, GTH, MBN, EJH, OJS, and AS acknowledge support from the European Research Council (ERC) under the European Union’s Horizon 2020 research and innovation programme (CartographY G.A. n. 804752). MNL acknowledges support from the ESA PRODEX programme. OJS acknowledges the support of the Science and Technology Facilities Council (STFC). RAG acknowledges the CNES PLATO and GOLF grants. This paper includes data collected by the Kepler and TESS missions and obtained from the MAST data archive at STScI. Funding for the Kepler mission is provided by the NASA Science Mission Directorate. Funding for the TESS mission is provided by the NASA Explorer Program. The computations described in this paper were performed using the University of Birmingham's BlueBEAR HPC service, which provides a High Performance Computing service to the University's research community. See \url{http://www.birmingham.ac.uk/bear} for more details.
 
\vspace{5mm}
\facilities{\texttt{Kepler}, TESS}

\software{\texttt{Python} \citep{vanRossum1991}, 
          \texttt{Matplotlib} \citep{hunter2007},
          \texttt{NumPy} \citep{Harris2020},
          \texttt{SciPy} \citep{Virtanen2020},
          \texttt{pandas} \citep{Reback2020},
          \texttt{Astropy} \citep{astropy2018},  
          \texttt{JAX} \citep{jax2018},
          \texttt{dynesty} \citep{Speagle2020}, 
          \texttt{emcee} \citep{Foreman-Mackey2013},
          \texttt{Echelle} (\url{https://github.com/danhey/echelle}),
          }

\bibliography{main}{}
\bibliographystyle{aasjournal}

\appendix
Table \ref{tab:modeIDvariables} shows the complete list of model parameters used in the various stages of \pbjam. The model parameters are separated in part by pertaining to either the $\ell=2,0$ stage or the $\ell=1$ stage. However, those that describe the envelope rotation and inclination of the star are used in both stages. In addition, some parameters are common to all some or all variants of the models in a given stage.

\begin{table}[h!]
    \centering
    \begin{tabular}{cllll}
       Parameter & Stage & Model & Description & Unit\\
       \hline
       $\dnu$         & $\ell=2,0$ & Modes & Large frequency separation. & $\muHz$\\
       $\numax$       & $\ell=2,0$ & Modes & Frequency at maximum oscillation power. & $\muHz$\\
       $\eps_\mathrm{p}$       & $\ell=2,0$ & Modes & Phase offset of the p-modes. & - \\
       $\dotwo$       & $\ell=2,0$ & Modes & $\ell=0,2$ mean frequency difference. & $\muHz$\\
       $\alpha_\mathrm{p}$     & $\ell=2,0$ & Modes & Curvature of the p-modes with radial order. & - \\
       $W_\mathrm{E}$ & $\ell=2,0$ & Modes & Envelope width. & $\muHz$\\ 
       $H_\mathrm{E}$ & $\ell=2,0$ & Modes & Envelope height. & ppm$^2/\muHz$ \\
       $\Gamma$       & $\ell=2,0$ & Modes & Mean mode width. & $\muHz$\\
       $\Teff$        & $\ell=2,0$ & Modes & Stellar effective temperature. & K\\
       $\bprp$        & $\ell=2,0$ & Modes & Gaia $\bprp$ color. & mag\\

       $a$ & $\ell=2,0$ & Background & Power of the high and mid-frequency Harvey profile. & ppm\\
       $b_1$ & $\ell=2,0$ & Background & Frequency of the high-frequency Harvey profile. & $\muHz$\\
       $c_1$ & $\ell=2,0$ & Background & Exponent of the high-frequency Harvey law profile. & - \\
       $b_2$ & $\ell=2,0$ & Background & Frequency of the mid-frequency Harvey profile. & $\muHz$\\
       $c_2$ & $\ell=2,0$ & Background & Exponent of the mid-frequency Harvey profile. & - \\
       $a_\mathrm{I}^{\dagger}$ & $\ell=2,0$ & Background & Power of the low-frequency Harvey profile. & ppm\\
       $b_\mathrm{I}^{\dagger}$ & $\ell=2,0$ & Background & Frequency of the low-frequency Harvey profile. & $\muHz$\\
       $c_\mathrm{I}^{\dagger}$ & $\ell=2,0$ & Background & Exponent of the low-frequency Harvey profile. & - \\
       $S^{\dagger}$ & $\ell=2,0$ & Background & Shot noise level. & ppm$^2/\muHz$\\
       
       $\delta\nu_{01}$ & $\ell=1$ &  MS, SG, RG & $\ell=0,1$ mean frequency difference. & $\muHz$\\
       $\DPione$ & $\ell=1$  & SG, RG & Period spacing of the $\ell=1$ modes. & s \\
       $\eps_\mathrm{g}$ & $\ell=1$  & SG, RG & Phase offset of the asymptotic g-modes. & - \\ 
       $p_{\mathrm{L}}$ & $\ell=1$  & SG & Coefficient for computing coupling matrix $L$. & - \\
       $p_{\mathrm{L}}$ & $\ell=1$  & SG & Coefficient for computing coupling matrix $D$. & - \\
       $q$ & $\ell=1$  & RG & Mixed mode coupling strength. & - \\
       
       $\delta\nu_{\mathrm{core}}^{\dagger}$ & $\ell=1$  & MS, SG, RG & Envelope rotation rate. & $\muHz$\\
       $\delta\nu_{\mathrm{env}}^{\dagger}$ & $\ell=2,0$, $\ell=1$ & All models & Core rotation rate. & $\muHz$\\
       $i_{\star}^{\dagger}$ &  $\ell=2,0$, $\ell=1$ & All models & Angle of the stellar rotation axis. & rad\\
    \end{tabular}
     
    \caption{List of parameters included in the mode identification models. The first column shows the parameter name, the second column denotes which stage of \pbjam uses the parameter. The parameters used in the $\ell=2,0$ stage are divided into those pertaining to describing the mode power density and those used in the background terms, where all the parameters are sampled simultaneously as part of the same spectrum model for this stage. In the $\ell=1$ stage there is a choice of the \MS, \SG, and \RG models, where each uses a combination of the listed parameters. The envelope rotation rate, $\delta\nu_{\mathrm{env}}$ and inclination angle $i_{\star}$ are used in both stages, and all the models. Parameters labeled with $\dagger$ are not included in the dimensionality reduction.}
    \label{tab:modeIDvariables}
\end{table}

\end{document}